\documentclass[11pt]{article}
\usepackage[margin=1in]{geometry}
\pdfoutput=1

\usepackage[ruled]{algorithm2e} 

\usepackage{standalone}
\usepackage{mathtools}
\usepackage{amsmath}
\usepackage{amsthm}
\usepackage{amsfonts}
\usepackage{natbib}
\usepackage{subfigure}
\usepackage{tikz}
\usetikzlibrary{fit,arrows,calc,positioning,shapes.geometric}
\usepackage{tikz-qtree}
\usepackage{soul}
\usepackage{enumerate}
\usepackage{graphicx}
\DeclareRobustCommand*\cal{\relax\mathcal}

\usepackage{thmtools}
\usepackage{thm-restate}
\theoremstyle{plain}
\newtheorem{theorem}{\protect\theoremname}
\theoremstyle{definition}
\newtheorem{definition}{\protect\definitionname}
\theoremstyle{plain}
\newtheorem{lemma}{\protect\lemmaname}
\theoremstyle{plain}
\newtheorem{corollary}{\protect\corollaryname}
\theoremstyle{plain}

\theoremstyle{plain}
\newtheorem{example}{\protect\examplename}
\providecommand{\definitionname}{Definition}
\providecommand{\lemmaname}{Lemma}
\providecommand{\corollaryname}{Corollary}
\providecommand{\propositionname}{Proposition}
\providecommand{\examplename}{Example}
\providecommand{\theoremname}{Theorem}

\def\mybeq{\begin{equation}\begin{array}{c}}
\def\myeeq{\end{array}\end{equation}}

\def\epsilonsad{{\epsilon_{\hbox{\scriptsize\rm sad}}}}

\def\Prox{{\hbox{\rm Prox}}}

\def\argmin{\mathop{\hbox{\rm argmin}}}
\def\nonrootsimplexes{\widehat{S}_Q}
\newtheorem{fact}{Fact}
\newtheorem{remark}{Remark}
\def\e{\ensuremath{{\epsilon}}}

\newcommand{\bbI}{\ensuremath{\mathbb{I}}}

\newcommand{\cD}{\ensuremath{\mathcal{D}}}

\newcommand{\cI}{\ensuremath{\mathcal{I}}}
\newcommand{\cX}{\ensuremath{\mathcal{X}}}
\newcommand{\cY}{\ensuremath{\mathcal{Y}}}
\newcommand{\cZ}{\ensuremath{\mathcal{Z}}}

\newcommand{\bea}{\begin{eqnarray}}
\newcommand{\eea}[1]{\label{#1}\end{eqnarray}}

\newcommand{\ese}{\end{eqnarray*}}
\newcommand{\bse}{\begin{eqnarray*}}
\def\beq{\begin{equation}}
\def\eeq{\end{equation}}

\def\fnote#1{\footnote}
\newcommand{\epr}{\hfill\hbox{\hskip 4pt \vrule width 5pt height 6pt depth 1.5pt}\vspace{0.5cm}\par}

\def\ra{\rangle}
\def\la{\langle}

\def\N{{\mathbb{N}}}

\def\R{{\mathbb{R}}}

\def\bE{{\mathbf{E}}}

\def\cD{{\cal D}}

\def\cS{{\cal S}}

\def\cX{{\cal X}}
\def\cY{{\cal Y}}

\def\Opt{{\hbox{\rm Opt}}}

\def\argmin{\mathop{\rm argmin}}

\def\ri{{\mathop{\rm ri}\,}}

\def\log{\mathop{{\rm log}}}

\newcommand{\wrt}{w.r.t.}

\newcommand{\cfr}{CFR}
\newcommand{\cfrp}{CFR+}
\newcommand{\mccfr}{MCCFR}
\newcommand{\egt}{EGT}

\newcommand{\mprox}{MP}

\newcommand{\dgf}{DGF}
\newcommand{\efg}{EFG}
\newcommand{\fom}{FOM}
\newcommand{\bspp}{BSPP}

\author{
  Christian Kroer\thanks{Supported by a Facebook Fellowship, the NSF under grants IIS-1617590, IIS-1320620, and IIS-1546752, and the ARO under awards
  W911NF-16-1-0061 and W911NF-17-1-0082.}\\
  Computer Science Department\\
  Carnegie Mellon University\\
  \tt{ckroer@cs.cmu.edu}
  \and
  Kevin Waugh\\
  Department of Computing Science\\
  University of Alberta\\
  \tt{kevin.waugh@gmail.com}
  \and
  Fatma K{\i}l{\i}n\c{c}-Karzan\thanks{Supported by NSF grant CMMI 1454548.}\\
  Tepper School of Business\\
  Carnegie Mellon University\\
  \tt{fkilinc@andrew.cmu.edu}
  \and
  Tuomas Sandholm\thanks{Supported by the NSF under grants
  IIS-1617590, IIS-1320620, and IIS-1546752, and the ARO under awards
  W911NF-16-1-0061 and W911NF-17-1-0082.}\\
  Computer Science Department\\
  Carnegie Mellon University\\
  \tt{sandholm@cs.cmu.edu}
}

\title{Theoretical and Practical Advances on Smoothing for Extensive-Form Games}
\begin{document}

\maketitle

\begin{abstract}
  Sparse iterative methods, in particular first-order methods, are known to be
among the most effective in solving large-scale two-player zero-sum
extensive-form games. The convergence rates of these methods depend heavily on
the properties of the distance-generating function that they are based on. We
investigate the acceleration of first-order methods for solving extensive-form
games through better design of the dilated entropy function---a class of
distance-generating functions related to the domains associated with the
extensive-form games. By introducing a new weighting scheme for the dilated
entropy function, we develop the first distance-generating function for the
strategy spaces of sequential games that only a logarithmic dependence on the
branching factor of the player. This result improves the convergence rate of
several first-order methods by a factor of $\Omega(b^dd)$, where $b$ is the
branching factor of the player, and $d$ is the depth of the game tree.

Thus far, counterfactual regret minimization methods have been faster in
practice, and more popular, than first-order methods despite their
theoretically inferior convergence rates. Using our new weighting scheme and
practical tuning we show that, for the first time, the excessive gap technique
can be made faster than the fastest counterfactual regret minimization
algorithm, {\cfrp}, in practice.
\end{abstract}

\section{Introduction}

Extensive-form games ({\efg}s) are a broad class of games; they model
sequential interaction, imperfect information, and outcome uncertainty. Nash
equilibria prescribe a particular notion of rational behavior in such games. In
the specific case of two-player zero-sum {\efg}s with perfect recall, an exact
Nash equilibrium can be computed in polynomial time using a Linear Program (LP)
whose size is linear in the size of the game tree~\citep{Stengel96:Efficient}.
However, in practice the LP approach has two major drawbacks limiting its applicability. First, the LP may be prohibitively large and may not fit in memory. Second, even when it does, the iterations of interior-point methods or the simplex algorithm are prohibitively expensive~\citep{Sandholm10:State}.
%
Practical methods for {\efg} solving tackle this issue through two complementary approaches: Abstraction and iterative game solvers with low memory requirements~\citep{Sandholm10:State}. In this paper we focus on the second approach. Iterative game solvers mainly fall in two categories: (i) counterfactual-regret-based methods~\citep{Zinkevich07:Regret,Lanctot09:Monte} achieving a convergence rate on the order of $O({1\over\epsilon^2})$, and (ii) first-order methods (FOMs)~\citep{Hoda10:Smoothing,Kroer15:Faster} achieving a convergence rate of $O({1\over\epsilon})$. The better convergence rate of {\fom}s makes them more attractive from a theoretical viewpoint. 
This paper investigates the acceleration of such {\fom}s for {\efg}s,  from both a theoretical and a numerical perspective.

Nash equilibrium computation of a two-player zero-sum \efg\ with perfect recall admits a Bilinear Saddle Point Problem ({\bspp}) formulation where the domains are given by the polytopes that encode strategy spaces of the players. The most efficient {\fom}s are designed to solve this \bspp. 
The classical {\fom}s to solve  {\bspp}s such as {\it mirror prox} (\mprox)~\citep{Nemirovski04:Prox} or the {\it excessive gap technique} (\egt)~\citep{Nesterov05:Excessive} utilize {\it distance-generating functions} ({\dgf}s)  to measure appropriate notions of distances over the domains. 
Then the convergence rate of these {\fom}s relies on the {\dgf}s and their relation to the domains 
in three critical ways: Through the strong convexity parameters of the {\dgf}s, the norm associated with the strong convexity parameter, and set widths of the domains as measured by the {\dgf}s. 

\citet{Hoda10:Smoothing} introduced a general framework for constructing {\dgf}s
for \emph{treeplexes}---a class of convex polytopes that generalize the domains
associated with the strategy spaces of an {\efg}. While they also established
bounds on the strong convexity parameter for their {\dgf}s in some special
cases, these lead to very weak bounds and result in slow convergence rates.
\citet{Kroer15:Faster} developed explicit strong convexity-parameter bounds for
entropy-based {\dgf}s (a particular subclass of {\dgf}s) for general {\efg}s,
and improved the bounds for the special cases considered by
\citet{Hoda10:Smoothing}. These bounds from~\citet{Kroer15:Faster} generate the
current state-of-the-art parameters associated with the convergence rate for
{\fom}s with $O({1\over \epsilon})$ convergence.

In this paper we construct a new weighting scheme for such entropy-based
{\dgf}s. This weighting scheme leads to new and improved bounds on the strong
convexity parameter associated with general treeplex domains. In particular, our
new bounds are first-of-their kind as they have no dependence on the branching
operation of the treeplex. Informally, our strong convexity result allows us to
improve the convergence rate of {\fom}s by a factor of $\Omega(b^{d}d)$ (where
$b$ is the average branching factor for a player and $d$ is the depth of the
\efg) compared to the prior state-of-the-art results from
\citet{Kroer15:Faster}. Our bounds parallel the simplex case for matrix games
where the entropy function achieves a logarithmic dependence on the
dimension of the simplex domain.


Finally, we complement our theoretical results with numerical experiments to
investigate the speed up of {\fom}s with convergence rate $O({1\over \e})$ and
compare the performance of these algorithms with the premier regret-based
methods \cfr\ and {\cfrp}~\citep{Tammelin15:Solving}. {\cfrp} is the fastest prior algorithm for computing Nash equilibria in {\efg}s when the entire tree can be traversed (rather than sampled).
\citet{Bowling15:Heads-up} used it to essentially solve the game limit Texas
hold'em.

CFR+ is also the algorithm used to accurately solve endgames in the Libratus
agent, which showed superhuman performance against a team of top Heads-Up
No-Limit Texas hold'em poker specialist professional players in the Brains vs AI
event~\footnote{Confirmed through author communication}. A slight variation\footnote{This variation was chosen for implementation
reasons, though, and has inferior practical iteration complexity.} of CFR+ was
used in the DeepStack agent \citet{Moravvcik17:Deepstack}, which beat a group of
professional players.
Our experiments show that {\fom}s are substantially
faster than both {\cfr} algorithms when using a practically tuned variant of our
{\dgf}. We also test the impact of stronger bounds on the strong convexity
parameter: we instantiate {\egt} with the parameters developed in this paper,
and compare the performance to the parameters developed by
\citet{Kroer15:Faster}. These experiments illustrate that the tighter parameters
developed here lead to better practical convergence rate.

The rest of the paper is organized as follows. Section~\ref{sec:related_work}
discusses related research. We present the general class of problems that we
address---bilinear saddle-point problems---and describe how they relate to
{\efg}s in Section~\ref{sec:bspp}. Then Section~\ref{sec:assumptions_and_setup}
describes our optimization framework. Section~\ref{sec:treeplexes} introduces
treeplexes, the class of convex polytopes that define our domains of the
optimization problems. Our focus is on dilated entropy-based {\dgf}s; we
introduce these in Section~\ref{sec:dgf} and present our main results---bounds
on the associated strong convexity parameter and treeplex diameter. In
Section~\ref{sec:mp_for_efg} we demonstrate the use of our results on
instantiating {\egt}. We compare our approach with the current state-of-art in
{\efg} solving and discuss the extent of theoretical improvements achievable via
our approach in Section~\ref{sec:theoretical_improvement}.
Section~\ref{sec:experiments} presents numerical experiments testing the effect
of various parameters on the performance of our approach as well as comparing
the performance of our approach to {\cfr} and {\cfrp}. We close with a summary
of our results and a few compelling further research directions in
Section~\ref{sec:conclusions}.

\section{Related work} \label{sec:related_work}

Nash equilibrium computation has received extensive attention in the
literature~\citep{Littman03:Polynomial,Lipton03:Playing,Gilpin07:Lossless,Zinkevich07:Regret,Daskalakis09:Complexity,Jiang11:Polynomial,Kroer14:Extensive-Form,Daskalakis15:Near}.
The equilibrium-finding problems vary quite a bit based on their
characteristics; here we restrict our attention to two-player zero-sum
sequential games.

\citet{Koller96:Efficient} present an LP whose size is linear in the size of the
game tree. This approach, coupled with lossless abstraction techniques, was used
to solve Rhode-Island hold'em~\citep{Shi01:Abstraction,Gilpin07:Lossless},
a game with $3.1$ billion nodes (roughly size $5\cdot10^7$ after lossless
abstraction). However, for games larger than this, the resulting LPs tend to not
fit in the computer memory thus requiring approximate solution techniques. These
techniques fall into two categories: iterative \e-Nash equilibrium-finding
algorithms and game abstraction techniques~\citep{Sandholm10:State}.

The most popular iterative Nash equilibrium algorithm is the
counterfactual-regret-minimization framework instantiated with regret matching
(\cfr)~\citep{Zinkevich07:Regret}, its sampling-based variant monte-carlo \cfr\
(MCCFR)~\citep{Lanctot09:Monte}, and \cfr\ instantitated with a new regret
minimization technique called regret matching plus ({\cfrp}). These
regret-minimization algorithms perform local regret-based updates at each
information set. Despite their slow convergence rate of $O(\frac{1}{\e^2})$,
they perform very well in pratice, especially {\cfrp}.
Recently,~\citet{Waugh15:Unified} showed, with some caveats, an interpretation
of \cfr\ as a \fom\ with $O(\frac{1}{\e^2})$ rate. Nonetheless, in this paper we
make a distinction between regret-based methods and $O(\frac{1}{\e})$ {\fom}s
for ease of exposition.

\citet{Hoda10:Smoothing} initially proposed {\dgf}s for {\efg}s leading to
$O(\frac{1}{\e})$ convergence rate when used with {\egt}. \citet{Kroer15:Faster}
improved these result for the dilated entropy function. \citet{Gilpin12:First}
give an algorithm with convergence rate $O(\ln(\frac{1}{\e}))$. Their bound has
a dependence on a certain condition number of the payoff matrix, which is
difficult to estimate; and as a result they show a bound of $O(\frac{1}{\e})$
which is independent of the condition number. Detailed comparisons to all
three algorithms discussed here are given in
Section~\ref{sec:theoretical_improvement}.

Finally, \citet{Bosansky14:Exact} develop an iterative double-oracle algorithm for exact equilibrium computation. This algorithm only scales for games where it can identify an equilibrium of small support, and thus suffers from the same performance issues as the general LP approach.

In addition to equilibrium-finding algorithms, another central topic in large-scale game solving has been automated abstraction~\citep{Sandholm10:State,Sandholm15:Abstraction}. 
Initially, this was used mostly for information abstraction \citep{Gilpin07:Lossless,Shi01:Abstraction,Zinkevich07:Regret}. Lately,  action abstraction approaches have gained considerable interest~\citep{Hawkin11:Automated,Hawkin12:Using,Brown14:Regret,Kroer14:Extensive-Form,Kroer16:Extensive_Imperfect}. 
Sequential game abstraction approaches with solution quality bounds have also emerged for stochastic~\citep{Sandholm12:Lossy} and extensive-form~\citep{Lanctot12:no-regret,Kroer14:Extensive-Form,Kroer16:Extensive_Imperfect} games more recently.

\section{Problem setup} \label{sec:bspp}
Computing a Nash equilibrium in a two-player zero-sum \efg\ 
with perfect recall can be formulated as a Bilinear Saddle Point Problem ({\bspp}):
\begin{equation}\label{eq:sequence_form_objective}
\min_{x \in \cX} \max_{y \in \cY} \langle x, Ay \rangle = \max_{y \in \cY} \min_{x \in \cX} \langle x,Ay \rangle .
\end{equation}
This is known as the {\it sequence-form} formulation~\citep{Romanovskii62:Reduction,Koller96:Efficient,Stengel96:Efficient}.
In this formulation, $x$ and $y$ correspond to the nonnegative strategy vectors for players~$1$ and $2$ and the sets $\cX,\cY$ are convex polyhedral reformulations of the sequential strategy space of these players. Here $\cX,\cY$ are defined by the constraints $Ex=e,Fy=f$, where each row of $E,F$ encodes part of the sequential nature of the strategy vectors, the right hand-side vectors $e,f$ are $\left|\cI_1\right|,\left|\cI_2\right|$-dimensional vectors, 
and $\cI_i$ is the information sets for player~$i$. For a complete treatment of this formulation, see \citet{Stengel96:Efficient}. 

Our theoretical developments mainly exploit the treeplex domain structure and are independent of other structural assumptions resulting from {\efg}s. Therefore, we describe our results for general {\bspp}s. 
We follow the presentation and notation of \citet{Juditsky11:First_general,Juditsky11:First_structure} for {\bspp}s. For notation and presentation of treeplex structure, we follow \citet{Kroer15:Faster}.

\subsection{Basic notation}
We let $\la x,y\ra$ denote the standard inner product of vectors $x,y$. Given a vector $x\in\R^n$, we let $\|x\|_p$ denote its $\ell_p$ norm given by $\|x\|_p:=\left( \sum_{i=1}^n |x_i|^p \right)^{1/p}$ for $p\in[1,\infty)$ and $\|x\|_\infty:=\max_{i\in[n]} |x_i|$ for $p=\infty$.
Throughout this paper, we use Matlab notation to denote vector and matrices, i.e., $[x;y]$ denotes the concatenation of two column vectors $x$, $y$.  
For a given set $Q$, we let $\ri(Q)$ denote its relative interior. Given $n\in\N$, we denote the simplex $\Delta_n:=\{x\in\R^n_+:\;\sum_{i=1}^n x_i=1\}$. 

\section{Optimization setup} \label{sec:assumptions_and_setup}
In its most general form a {\bspp} is defined as
\[
\Opt:=\max_{y\in \cY}\min_{x\in \cX} \phi(x,y),
\eqno{(\cS)}
\]
where $\cX,\cY$ are nonempty convex compact sets in Euclidean spaces $\bE_x,\bE_y$ and $\phi(x,y)=\upsilon+\langle a_1,x\rangle + \langle a_2,y\rangle +\langle y,Ax\rangle$. We let $\cZ:=\cX\times \cY$; so $\phi(x,y):\cZ\to\R$. In the context of {\efg} solving, $\phi(x,y)$ is simply the inner product given in \eqref{eq:sequence_form_objective}.

The {\bspp} $(\cS)$ gives rise to two convex optimization problems that are dual to each other:
\begin{equation*}\label{neq1}
\begin{array}{rclcr}
\Opt(P)&=&\min_{x\in \cX}[ \overline{\phi}(x):=\max_{y\in \cY} \phi(x,y)]&&(P),\\
\Opt(D)&=&\max_{y\in \cY}[ \underline{\phi}(y):=\min_{x\in \cX} \phi(x,y)] &&(D),\\
\end{array}
\end{equation*}
with $\Opt(P)=\Opt(D)=\Opt$.
It is well known that the solutions to $(\cS)$ --- the saddle points of $\phi$ on $\cX\times \cY$ --- are exactly the pairs $z=[x;y]$ comprised of  optimal solutions to the problems $(P)$ and $(D)$. 
We quantify the accuracy of a candidate solution $z=[x;y]$ 
with the \emph{saddle point residual} 
\bse\label{epssad}
\epsilonsad(z)&:=&\overline{\phi}(x)-\underline{\phi}(y) 
=\underbrace{\left[\overline{\phi}(x)-\Opt(P)\right]}_{\geq0}+
\underbrace{\left[\Opt(D)-\underline{\phi}(y)\right]}_{\geq0}.
\ese
In the context of \efg, $\epsilonsad(z)$ measures the proximity to being an \e-Nash equilibrium.


\subsection{General framework for {\fom}s} \label{sec:mirror_prox}
Most {\fom}s capable of solving {\bspp} $(\cS)$ are quite flexible in terms of adjusting to the geometry of the problem characterized by the domains $\cX,\cY$ of the {\bspp} $(\cS)$. The following components are standard in forming the setup for such {\fom}s (we present components for $\cX$, analogous components are used for $\cY$):
\begin{itemize}
\item \emph{Vector norm}: $\|\cdot\|_\cX$ on the Euclidean space $\bE$ where the domain $\cX$ of $(\cS)$ lives, along with its dual norm $\|\zeta\|_\cX^*=\max\limits_{\|x\|_\cX\leq1}\langle\zeta,x\rangle$.
\item \emph{Matrix norm}: $\|A\|=\max_y\left\{ \|Ay\|_\cX^* : \|y\|_\cY=1 \right\}$ based on the vector norms $\|\cdot\|_\cX,\|\cdot\|_\cY$. 
\item \emph{Distance-Generating Function} ({\dgf}): A function $\omega_\cX(x):\cX\rightarrow \R$, which is convex and continuous on $\cX$, and admits a continuous selection of subgradients  $\omega_\cX'(x)$ on the set $\cX^\circ:=\{x\in \cX:\partial\omega_\cX(x)\neq\emptyset\}$ (here $\partial\omega_\cX(x)$ is a subdifferential of $\omega_\cX$ taken at $x$), and is strongly convex with modulus $\varphi_\cX$ {\wrt} the norm  $\|\cdot\|_\cX$:
    \begin{equation}
    \forall x',x''\in \cX^\circ:~ \langle \omega_\cX'(x')-\omega_\cX'(x''),x'-x''\rangle \geq \varphi_\cX \|x'-x''\|_\cX^2. \label{eq:strong_convexity_definition}
    \end{equation}
\item \emph{Bregman distance}: $V(u\|x):=\omega_\cX(u)-\omega_\cX(x)-\langle \omega_\cX'(x),u-x\rangle$ for all $x\in \cX^\circ$ and $u\in \cX$.
\item  \emph{Prox-mapping}: Given a \emph{prox center} $x\in \cX^\circ$,
  \[
  \Prox_x(\xi):=\argmin\limits_{u\in \cX}\left\{\langle \xi,u\rangle +V(u\|x)\right\}: \bE\to \cX^\circ.
  \]
For properly chosen stepsizes, the prox-mapping becomes a contraction.  This is critical in the convergence analysis of {\fom}s. Furthermore, when the {\dgf} is taken as the squared $\ell_2$ norm, the prox mapping becomes the usual projection operation of the vector $x-\xi$ onto $\cX$.
\item \emph{$\omega$-center}: $x_\omega:=\argmin\limits_{x\in \cX}\omega_\cX(x)\in \cX^\circ$ of $\cX$.
\item \emph{Set width}: $\Omega_x:=\max\limits_{x\in \cX}V(x\|{x_\omega})\leq\max\limits_{x\in \cX}\omega_\cX(x)-\min\limits_{x\in \cX}\omega_\cX(x)$.
\end{itemize}
 
The distance-generating functions $\omega_\cX,\omega_\cY$ can be used to create \emph{smoothed approximations} to $\overline{\phi},\underline{\phi}$ as follows~\citep{Nesterov05:Smooth}:
\begin{align}
  \overline{\phi}_{\mu_2}(x) = \max_{y\in \cY} \left\{\phi(x,y) - \mu_2\omega_\cY(y)\right\},\label{eq:smoothed_y}\\
  \underline{\phi}_{\mu_1}(y) = \min_{x\in \cX} \left\{\phi(x,y) + \mu_1\omega_\cX(x)\right\},  \label{eq:smoothed_x}
\end{align}
where $\mu_1,\mu_2>0$ are smoothness parameters denoting the amount of smoothing
applied. Let $y_{\mu_2}(x)$ and $x_{\mu_1}(y)$ refer to the $y$ and $x$ values attaining the optima in \eqref{eq:smoothed_y} and \eqref{eq:smoothed_x}. These can be thought of as \emph{smoothed best responses}. 
\citet{Nesterov05:Smooth} shows that the gradients of the functions $\overline{\phi}_{\mu_2}(x)$ and $\underline{\phi}_{\mu_1}(y)$ exist and are
Lipschitz continuous. The gradient operators and Lipschitz constants are given as follows
\begin{align*}
  &\nabla \overline{\phi}_{\mu_2}(x) = a_1 + Ay_{\mu_2}(x) \quad\text{and}\quad   \nabla \underline{\phi}_{\mu_1}(y) = a_2 + A^\top x_{\mu_1}(y), \\
  &L_1\left(\overline{\phi}_{\mu_2}\right) = \frac{\|A\|^2}{\varphi_\cY\mu_2}  \quad\text{and}\quad   L_2\left(\underline{\phi}_{\mu_1}\right) = \frac{\|A\|^2}{\varphi_\cX\mu_1}.
\end{align*}
Based on this setup, we formally state the
\emph{Excessive Gap Technique} ({\egt}) of \citet{Nesterov05:Excessive} in
Algorithm~\ref{alg:egt}.

 \begin{minipage}{0.49\textwidth}
  \begin{algorithm}[H]
   \caption{EGT}
   \label{alg:egt}
  \SetKwInOut{Input}{input}
  \SetKwInOut{Output}{output}
  \Input{$\omega$-center $z_\omega$, \dgf\ weights $\mu_1,\mu_2$, and $\epsilon>0$}
  \Output{$z^t(=[x^t;y^t])$} 
  $x^{0} = \Prox_{x_{\omega}}\left( \mu_1^{-1} \nabla \overline{\phi}_{\mu_2}(x_{\omega})\right)$\;
  $y^0 = y_{\mu_2}(x_{\omega})$\;
  $t = 0; z_1\coloneqq z_\omega$\;
  \While{$\epsilonsad(z^t)>\epsilon$}{
    {$\tau_t = \frac{2}{t+3}$\;}
    \uIf{$t$ is even}{
      $(\mu_1^{t+1},x^{t+1},y^{t+1}) = Step(\mu_1^{t}, \mu_2^t, x^t, y^t, \tau)$
    }
    \Else{
      $(\mu_2^{t+1},y^{t+1},x^{t+1}) = Step(\mu_2^t, \mu_1^t, y^t, x^t, \tau)$
    }
    {$t=t+1$\;}
  }
  \end{algorithm}
\end{minipage}%
\hfill
\begin{minipage}{0.39\textwidth}
  \begin{algorithm}[H]
    \caption{Step}
    \label{alg:step}
  \SetKwInOut{Input}{input}
  \SetKwInOut{Output}{output}
  \Input{$\mu_1,\mu_2, x, y, \tau$}
  \Output{$\mu_1^+,x_+,y_+$} 
  $\hat{x} = \left( 1-\tau \right)x + \tau x_{\mu_1}(y)$\;
  $y_+ = \left( 1-\tau \right)y + \tau y_{\mu_2}(\hat{x})$\;
  $\tilde{x} = \Prox_{x_{\mu_1}(y)}\left(\frac{\tau}{\left( 1-\tau \right)\mu_1} \nabla \overline{\phi}_{\mu_2}(\hat{x})\right)$\;
  $x_+ = \left( 1-\tau \right)x + \tau \tilde{x}$\;
  $\mu_1^+ = \left( 1-\tau \right)\mu_1$\;
  \end{algorithm}
\end{minipage}

The \egt\ algorithm alternates between taking steps focused on $\cX$ and $\cY$. Algorithm~\ref{alg:step} shows a single step focused on $\cX$. Steps focused on $y$ are completely analogous. Algorithm~\ref{alg:egt} shows how the alternating steps and stepsizes are computed, as well as how initial points are selected.

Suppose the initial values $\mu_1,\mu_2$ in the {\egt} algorithm satisfy
$\mu_1=\frac{\varphi_\cX}{L_1(\overline{\phi}_{\mu_2})}$. Then, at every iteration $t\geq 1$ of
the {\egt} algorithm, the corresponding solution $z^t=[x^t;y^t]$ satisfies
$x^t\in \cX$, $y^t\in \cY$, and
\[
\overline{\phi}(x^t)-\underline{\phi}(y^t) = \epsilonsad(z^t) \leq  \frac{4\|A\|}{T+1}\sqrt{\frac{\Omega_\cX\Omega_\cY}{\varphi_\cX\varphi_\cY}}.
\]
Consequently, \citep{Nesterov05:Excessive} proves that the {\egt} algorithm has a convergence rate of $O(\frac{1}{\e})$.

\section{Treeplexes}
\label{sec:treeplexes}
\citet{Hoda10:Smoothing} introduce the {\em treeplex}, a class of convex polytopes that encompass the sequence-form description of strategy spaces in perfect-recall {\efg}s. 
\begin{definition}
  Treeplexes are defined recursively:
  \begin{enumerate}
  \item {\em Basic sets}: The standard simplex $\Delta_m$ 
  is a treeplex.
  \item {\em Cartesian product}:  If $Q_1,\ldots, Q_k$ are treeplexes, then $Q_1 \times \cdots \times Q_k$ is a treeplex.
  \item {\em Branching}: Given a treeplex $P\subseteq \left[ 0,1 \right]^p$, a collection of treeplexes $Q=\left\{ Q_1,\ldots,Q_k \right\}$ where $Q_j\subseteq \left[ 0,1 \right]^{n_j}$, and $l=\left\{l_1,\ldots,l_k \right\} \subseteq \left\{ 1,\ldots, p \right\}$, the set defined by 
\[
P\framebox{l}Q \coloneqq \left\{ \left(u,q_1,\ldots,q_k\right) \in \R^{p+\sum_j n_j}  :~ u\in P,~ q_1\in u_{l_1} \cdot Q_1,\, \ldots, q_k\in u_{l_k} \cdot Q_k \right\}
\]
    is a treeplex. 
    In this setup, we say $u_{l_j}$ is the branching variable for the treeplex $Q_j$.
  \end{enumerate}
\end{definition}

A treeplex is a tree of simplexes where children are connected to their parents through the branching operation. In the branching operation, the child simplex domain is scaled by the value of the parent branching variable. Understanding the treeplex structure is crucial because the proofs of our main results rely on induction over these structures. For {\efg}s, the simplexes correspond to the information sets of a single player and the whole treeplex represents that player's strategy space.
The branching operation has a sequential interpretation: The vector $u$ represents the decision variables at certain stages, while the vectors $q_j$ represent the decision variables at the $k$ potential following stages, depending on external outcomes. Here $k\leq p$ since some variables in $u$ may not have subsequent decisions. For treeplexes, \citet{Stengel96:Efficient} has suggested a polyhedral representation of the form $Eu=e$ where the matrix $E$ has its entries from $\left\{ -1,0,1 \right\}$ and the vector $e$ has its entries in $\left\{0,1\right\}$. 

For a treeplex $Q$, we denote by $S_Q$ the index set of the set of simplexes contained in $Q$ (in an \efg\ $S_Q$ is the set of information sets belonging to the player). For each $j\in S_Q$, the treeplex rooted at the $j$-th simplex $\Delta^j$ is referred to as $Q_j$. Given vector $q\in Q$ and simplex $\Delta^j$, we let $\bbI_j$ denote the set of indices of $q$ that correspond to the variables in $\Delta^j$ and define $q^j$ to be the sub vector of $q$ corresponding to the variables in $\bbI_j$.  For each simplex $\Delta^j$ and branch $i\in \bbI_j$, the set $\cD_j^i$ represents the set of indices of simplexes reached immediately after $\Delta^j$ by taking branch $i$ (in an \efg\ $\cD_j^i$ is the set of potential next-step information sets for the player). Given a vector $q\in Q$,  simplex $\Delta^j$, and index $i\in \bbI_j$, each child simplex $\Delta^k$ for every $k\in \cD_j^i$ is scaled by $q_i$. Conversely, for a given simplex $\Delta^j$, we let $p_j$ denote the index in $q$ of the parent branching variable $q_{p_j}$ that $\Delta^j$ is scaled by. We use the convention that $q_{p_j}=1$ if $Q$ is such that no branching operation precedes $\Delta^j$. For each $j\in S_Q$, $d_j$ is the maximum depth of the treeplex rooted at $\Delta^j$, that is, the maximum number of simplexes reachable through a series of branching operations at $\Delta^j$. Then $d_Q$ gives the depth of $Q$. We use $b_Q^j$ to identify the number of branching operations preceding the $j$-th simplex in $Q$. We will say that a simplex $j$ such that $b_Q^j=0$ is a \emph{root simplex}.

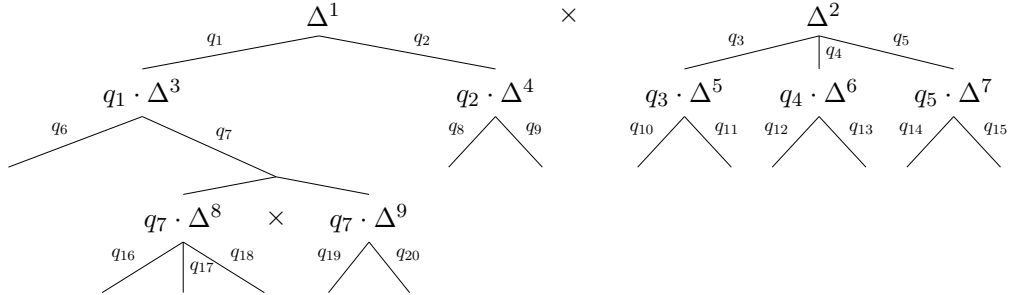
\begin{figure}[!h]
  \begin{center}
    \scalebox{0.95}{
      \begin{tikzpicture}
\tikzstyle{level 1}=[sibling distance=5pt]
\tikzstyle{level 2}=[sibling distance=30pt]
\tikzstyle{level 3}=[sibling distance=15pt, level distance = 20pt]
\tikzstyle{level 4}=[sibling distance=25pt]
\begin{scope}[xshift=2cm]
\Tree [.\node(a1){$\ \Delta^1$};
    \edge node[auto=right,scale=0.7]{$q_1$};
    [.{ $q_1\cdot\Delta^3$}
        \edge node[auto=right,scale=0.7]{$q_6$};
        {}
        \edge node[auto=left,scale=0.7]{$q_7$};
        [
        [.\node(d7){$q_7\cdot\Delta^8$};
            \edge node[auto=right,scale=0.7]{$q_{16}$};
            {}
            \edge node[auto=left,scale=0.7]{$q_{17}$}; 
            {}
            \edge node[auto=left,scale=0.7]{$q_{18}$}; 
            {}
            ]
        [.\node(d8){$q_7\cdot\Delta^9$};
            \edge node[auto=right,scale=0.7]{$q_{19}$};
            {}
            \edge node[auto=left,scale=0.7]{$q_{20}$}; 
            {}
            ]
        ]
     ]
    \edge node[auto=left,scale=0.7]{$q_2$};
    [. {$q_2\cdot\Delta^4$}
    \edge node[auto=right,scale=0.7]{$q_8$};
            {} 
    \edge node[auto=left,scale=0.7]{$q_9$};
            {} 
    ]
    ]
\end{scope}
\begin{scope}[xshift=9cm]
\Tree [.\node(a2){$\ \Delta^2$};
    \edge node[auto=right,scale=0.7]{$q_3$};
    [. {$q_3\cdot\Delta^5$}
    \edge node[auto=right,scale=0.7]{$q_{10}$};
            {} 
    \edge node[auto=left,scale=0.7]{$q_{11}$};
            {} 
    ]
    \edge node[auto=left,scale=0.7]{$q_{4}$};
    [. {$q_4\cdot\Delta^6$}
    \edge node[auto=right,scale=0.7]{$q_{12}$};
            {} 
    \edge node[auto=left,scale=0.7]{$q_{13}$};
            {} 
    ]
    \edge node[auto=left,scale=0.7]{$q_{5}$};
    [. {$q_5\cdot\Delta^7$}
    \edge node[auto=right,scale=0.7]{$q_{14}$};
            {} 
    \edge node[auto=left,scale=0.7]{$q_{15}$};
            {} 
    ]
]
\end{scope}
\path (a1) -- node {$\times$} (a2);
\path (d7) -- node {$\times$} (d8);
\end{tikzpicture}
    }
  \end{center}
  \caption{An example treeplex constructed from $9$ simplexes. Cartesian product operation is denoted by $\times$.  }
  \label{fig:treeplex}
\end{figure}

Figure~\ref{fig:treeplex} illustrates an example treeplex $Q$. $Q$ is
constructed from nine two-to-three-dimensional simplexes $\Delta^1,\ldots,
\Delta^9$. At level~$1$, we have two root simplexes, $\Delta^1,\Delta^2$,
obtained by a Cartesian product (denoted by $\times$). We have maximum depths $d_1=2$, $d_2=1$ beneath them.
Since there are no preceding branching operations, the parent variables for
these simplexes $\Delta^1$ and $\Delta^2$ are $q_{p_1}=q_{p_2}=1$. For
$\Delta^1$, the corresponding set of indices in the vector $q$ is
$\bbI_1=\left\{ 1,2 \right\}$, while for $\Delta^2$ we have $\bbI_2=\left\{ 3,4,
  5 \right\}$. At level~$2$, we have the simplexes $\Delta^3,\ldots,\Delta^7$.
The parent variable of $\Delta^3$ is $q_{p_3}=q_1$; therefore, $\Delta^3$ is
scaled by the parent variable $q_{p_3}$. Similarly, each of the simplexes
$\Delta^3,\ldots,\Delta^7$ is scaled by their parent variables $q_{p_j}$ that
the branching operation was performed on. So on for $\Delta^8$ and $\Delta^9$ as
well. The number of branching operations required to reach simplexes
$\Delta^1,\Delta^3$ and $\Delta^8$ is $b_Q^1=0,b_Q^3=1$ and $b_Q^8=2$,
respectively.

Note that we allow more than two-way branches; hence our formulation follows that of~\citet{Kroer15:Faster} and differs from that of \citet{Hoda10:Smoothing}.
As discussed in~\citet{Hoda10:Smoothing}, it is possible to model sequence-form games by treeplexes that use only two-way branches. Yet, this can cause a large increase in the depth of the treeplex, thus leading to significant degradation in the strong convexity parameter. Because we handle multi-way branches directly in our framework, our approach is more effective in taking into account the structure of the sequence-form game and thereby resulting in better bounds on the associated strong convexity parameters and thus overall convergence rates. 

Our analysis requires a measure of the size of a treeplex $Q$. Thus, we define
$M_Q\coloneqq \max_{q\in Q} \|q\|_1$. 

In the context of {\efg}s, suppose $Q$ encodes player 1's strategy space;  then $M_Q$ is the maximum number of information sets with nonzero probability of being reached when player 1 has to follow a pure strategy  while the other player may follow a mixed strategy. 
We also let 
\begin{equation}\label{eq:max_norm_cutoff}
M_{Q,r}\coloneqq \max_{q\in Q} \sum_{j\in S_Q: b_Q^j \leq r} \|q^j\|_1. 
\end{equation}
Intuitively, $M_{Q,r}$ gives the maximum value of the $\ell_1$ norm of any vector $q\in Q$ after removing the variables corresponding to simplexes that are not within $r$ branching operations of the root of $Q$.

\begin{example} \label{ex:square_root_m}
In order to illustrate $M_Q$ and compare it to the size of $|S_Q|$, let us now
consider an example of an EFG and its corresponding treeplexes. Consider a game
where two players take turns choosing among $k$ actions, and each player chooses
actions $d$ times before leaf nodes are reached. In the treeplex $Q$ of Player
$1$, each time Player $1$ chooses among $k$ actions constitutes a size $k$
branching operation, and every time Player $2$ chooses among $k$ actions
constitutes a size $k$ Cartesian product operation. The total dimensionality of
the treeplex, $|S_Q|$, is $k^{2d}$, while the value of $M_Q$ is $k^d$ (since
only Cartesian products blow up). Thus, $M_Q$ is square root of the size of
$|S_Q|$.
\end{example}

\section{Dilated entropy functions with bounded strong convexity}
\label{sec:dgf}
In this section we introduce {\dgf}s for domains with treeplex structures and establish their strong convexity parameters with respect to a given norm (see \eqref{eq:strong_convexity_definition}). 

The basic building block in our construction is the {\em entropy} {\dgf} given by $\omega_e(z)=\sum_{i=1}^n z_i\log(z_i),$
for the simplex $\Delta_n$. 
It is well-known that $\omega_e(\cdot)$ is strongly convex with modulus~$1$ with respect to the $\ell_1$ norm on $\Delta_n$ (see~\citet{Juditsky11:First_general}). We will show that a suitable modification of this function achieves a desirable strong convexity parameter for the  treeplex domain.

The treeplex structure is naturally related to the {\em dilation operation}~\citep{Hiriart01:Fundamentals} defined as follows: Given a compact set $K\subseteq \R^d$ and a function $f: K \rightarrow \R$, we first define 
\[
\bar{K} \coloneqq \left\{ (t,z) \in \R^{d+1} :~ t \in \left[ 0,1 \right],\; z \in t \cdot K \right\}.
\]
\begin{definition} Given a function $f(z)$, the {\em dilation operation} is the function $\bar{f} : \bar{K} \rightarrow \R$ given by
\[
\bar{f}(z,t) = \begin{cases}
  t\cdot f(z/t) & \mbox{if $t > 0$} \\
  0 & \mbox{if $t = 0$} \\
\end{cases}.
\]
\label{def:dilation_operation}
\end{definition}

The dilation operation preserves convexity, and thus we define the following
convex function by dilating the entropy function over the simplexes of a
treeplex:
\begin{definition}\label{def:dilated_entropy_function}
  Given a treeplex $Q$ and weights $\beta_j>0$ for each $j \in S_Q$, we define the {\em dilated entropy function} as
  \[
  \omega(q)=\sum_{j \in S_Q} \beta_j\sum_{i\in \bbI_j} q_{i}\log\frac{q_i}{q_{p_j}} \quad\mbox{for any }q\in Q,
  \]
  where we follow the treeplex notation and $p_j$ is the index of the branching variable preceding $\Delta^j$, with the convention that $q_{p_j}=1$ if $\Delta^j$ has no branching operation preceding it.
\end{definition}

\begin{remark}
  Note that the dilated entropy function $\omega(\cdot)$ defined above is twice
  differentiable in the relative interior of treeplex $Q$ and admits a
  continuous gradient selection. Moreover, for weights $\beta_j$ that scale
  appropriately with depth $d_j$, we will demonstrate that it is strongly convex
  {\wrt} the $\ell_1$ norm. Thus, the dilated entropy function is compatible
  with the $\ell_1$ norm, as required by the {\bspp} setup. \epr
\end{remark}

We would also like the prox-mapping associated with our DGF to be efficiently
computable. \citet{Hoda10:Smoothing} show that for any dilated function, its
prox operator on a treeplex can be easily computed through a recursive bottom-up
traversal involving the prox mappings associated with the function being dilated
on individual simplexes. Since the entropy prox function can be computed in
closed form on a simplex, the dilated entropy function can be computed by a
single treeplex traversal involving closed-form expressions on each simplex.

Definition~\ref{def:dilated_entropy_function} above leads to 
a subset of the {\dgf}s considered by \citet{Hoda10:Smoothing}. Our main theoretical result shows that by a careful selection of the weights $\beta_j$, we can significantly improve the strong convexity bounds associated with the dilated entropy function. 
We will consider weights that satisfy the following recurrence:
\begin{equation}
  \label{eq:weight_recurrence}
  \begin{aligned}
    \alpha_j &= 1 + \max_{i\in \bbI_j}\sum_{k \in \cD^i_j} \frac{\alpha_k\beta_k}{\beta_k - \alpha_k},&\quad \forall j \in S_Q,\\
    \beta_j &> \alpha_j,&\quad \forall i\in \bbI_j \mbox{ and } \forall j \in S_Q ~\text{s.t.}~ b_Q^j > 0,\\
    \beta_j &= \alpha_j,&\quad \forall i\in \bbI_j \mbox{ and } \forall j \in S_Q ~\text{s.t.}~ b_Q^j = 0.
  \end{aligned}
\end{equation}
Intuitively, $\alpha_j$ represents the negative terms that the weight $\beta_j$
has to cancel out: the constant $1$ represents the negative term resulting from the squared norm in the strong convexity requirement; the summation term represents the amount of negative terms accumulated from the induction on simplexes
descending from simplex $j$. The qualifications on $\beta_j$ ensure that
$\beta_j$ is set such that it at least cancels out the negative terms; the
difference $\beta_j-\alpha_j$ controls the amount of negative value the parent
simplex has to make up. This is why we set $\beta_j=\alpha_j$ when $b_Q^j=0$. As part
of the proof of Lemma~\ref{le:strong_convexity_rhs_sum} we will see why we
require a strict inequality $\beta_j>\alpha_j$ for non-root simplexes.

Based on recurrence~\eqref{eq:weight_recurrence}, our main results establish  strong convexity of our dilated entropy {\dgf} {\wrt} the $\ell_2$ and $\ell_1$ norms:
\begin{restatable}[]{theorem}{strongconvexityltwo}
  \label{th:recurrence_bound_l2}
  For a treeplex $Q$, the dilated entropy function with weights satisfying recurrence (\ref{eq:weight_recurrence}) is strongly convex  with modulus $1$ with respect to the $\ell_2$ norm.
\end{restatable}
\begin{restatable}[]{theorem}{strongconvexitylone}
  \label{th:recurrence_bound_l1}
  For a treeplex $Q$, the dilated entropy function with weights satisfying recurrence (\ref{eq:weight_recurrence}) is strongly convex  with modulus $\frac{1}{M_Q}$ with respect to the $\ell_1$ norm.
\end{restatable}
We give the proofs of Theorems~\ref{th:recurrence_bound_l2} and~\ref{th:recurrence_bound_l1} in Section~\ref{sec:strong_convexity_proof}. Based on Theorem~\ref{th:recurrence_bound_l1}, we get the following corollary:
\begin{corollary}\label{co:strong_convexity_dilated_entropy}
  For a treeplex $Q$, the dilated entropy function with weights $\beta_{j}=2+
  \sum_{r=1}^{d_j}2^{r}(M_{Q_{j},r}-1)$ for all $j \in S_Q$ is strongly convex
  with modulus $\frac{1}{M_{Q}}$ {\wrt} the $\ell_{1}$ norm.
\end{corollary}
Corollary~\ref{co:strong_convexity_dilated_entropy} follows easily from
Theorem~\ref{th:recurrence_bound_l1} and a recursive interpretation of the weights, which is presented as Fact~\ref{fact:Beta} in the next section. In particular, a specific choice of weights in Fact~\ref{fact:Beta} immediately satisfies the recurrence~\eqref{eq:weight_recurrence} and leads to Corollary~\ref{co:strong_convexity_dilated_entropy}.

To our knowledge, the best strong convexity bounds for general treeplexes were
proved in \citet{Kroer15:Faster}. Using weights $\beta_{j}=2^{d_j}M_{Q_j}$ they
show strong convexity modulus $\frac{1}{|S_Q|}$ {\wrt} the $\ell_{1}$ norm.
Corollary~\ref{co:strong_convexity_dilated_entropy} improves the prior bounds by
exchanging a factor of $\left|S_Q\right|$ with a factor of $M_Q$. Note that
$\left|S_Q\right|$ is tied to the branching factor associated with branching
operations in the treeplex $Q$ whereas $M_Q$ is not. Thus, our result
removes the dependence of the strong convexity parameter on the branching
factor and hence significantly improves upon \citet{Kroer15:Faster}.

In Theorem~\ref{the:entropy_diameter} we use our strong convexity result to establish a polytope diameter that has only a logarithmic dependence on the branching factor. As a consequence, the associated dilated entropy \dgf\ when used in {\fom}s such as \mprox\ and \egt\ for solving {\efg}s leads to the same improvement in their convergence rate.


\subsection{Preliminary results for the proofs of
  our main results}
\label{sec:basic_facts}
We start with some simple facts and a few technical lemmas that are used in our proofs.


\begin{fact}\label{fact:M_Q} 
Given a treeplex $Q$, we have, respectively, for all $i\in\mathbb{I}_{j},  j\in S_Q$ and all $d=1,\ldots, d_Q, q\in Q$:
\begin{align*}
&(a)\quad M_{Q_{j}} \geq 1+\sum_{l\in\mathcal{D}_{{j}}^{i}}M_{Q_{l}}, &  
(b)\quad M_{Q} \geq \sum_{j\in S_Q: d_j=d}q_{p_j} M_{Q_j}.
\end{align*}
\end{fact}
\begin{proof}
  The first inequality was established in \citet[Lemma 5.7]{Kroer15:Faster}. The
  second follows by using $M_Q = \sum_{j} q_i$ for some $q$, and inductively
  replacing terms belonging to simplexes j at the bottom with $M_{Q_j}$. The
  result follows because branching operations cancel out by summing to $1$.
\end{proof}

Our next observation follows from Fact~\ref{fact:M_Q}(a) and is advantageous in suggesting a practically useful choice of the weights $\beta_j$ that can be used for Theorem~\ref{th:recurrence_bound_l1} to arrive at Corollary~\ref{co:strong_convexity_dilated_entropy}.
\begin{fact}\label{fact:Beta}
Let $Q$ be a treeplex and $\beta_{j}=2+ \sum_{r=1}^{d_j}2^{r}(M_{Q_{j},r}-1)$ for all $j \in S_Q$ as in Corollary~\ref{co:strong_convexity_dilated_entropy}. Then Fact~\ref{fact:M_Q}(a) 
implies
  $\beta_j \geq 2 + \sum_{k \in \cD^i_j} 2\beta_k, \forall i\in \bbI_j \mbox{ and } \forall j \in S_Q.$
\end{fact}
Consequently, by selecting $\beta_j=2\alpha_j,$ and $\alpha_j=1 + \sum_{r=1}^{d_j}2^{r-1}(M_{Q_{j},r}-1)$ for all $i\in \bbI_j$ and for all $j \in S_Q$ such that $b_Q^j > 0$, we immediately satisfy the conditions of the recurrence in \eqref{eq:weight_recurrence}. 

Given a twice differentiable function $f$, we let $\nabla^2f(z)$ denote its Hessian at $z$. Our analysis is based on the following sufficient condition for strong convexity of a twice differentiable function:
\begin{fact} \label{fac:strong_convexity_hessian}
  A twice-differentiable function $f$ is strongly convex with modulus $\varphi$ with respect to a norm $\|\cdot\|$ on nonempty convex set $C\subset\R^n$ if 
$
  h^\top \nabla^2f(z) h \geq \varphi\|h\|^2,\ \forall h\in\R^n, z\in C^\circ.
$
\end{fact} 

For simplexes $\Delta^j$ at depth~$1$, there is no preceding branching operation; so the variables $h_{p_j},q_{p_j}$ do not exist. We circumvent this with the convention  $h_{p_j}=0,q_{p_j}=1$ for such $j\in S_Q$.

In our proofs we will use the following expression for $ h^\top \nabla^2\omega(q) h$. 
\begin{lemma}\label{lem:hessian}
Given a  treeplex $Q$ and a dilated entropy function $\omega(\cdot)$ with weights $\beta_j>0$,  we have
\begin{equation}\label{eq:hessian_quadratic_count_at_parent}
     h^\top\nabla^2\omega(q)h = \sum_{j\in S_Q} \beta_j \left[ \sum_{i\in \bbI_j}\left( \frac{h_i^2}{q_i}  -  \frac{2h_ih_{p_j}}{q_{p_j}} \right)   + \frac{h_{p_j}^2}{q_{p_j}}\right] \quad\forall q\in \ri(Q) \mbox{ and } \forall h\in\R^n.
\end{equation}
\end{lemma}
We provide the proof of Lemma~\ref{lem:hessian} in the appendix. It simply follows from taking the second-order partial derivatives and rearranging terms.

\subsection{Proofs of our main theorems}
\label{sec:strong_convexity_proof}

The majority of the work for our strong-convexity results is performed by the following lemma, from which our strong convexity results follow easily.
\begin{lemma}
  \label{le:strong_convexity_rhs_sum}
  For any treeplex $Q$, the dilated entropy function with weights satisfying recurrence \eqref{eq:weight_recurrence} satisfies the following inequality:
\begin{equation}\label{eq:strong_convexity_rhs_sum}
  h^\top\nabla^{2}\omega(q)h\geq\sum_{j\in S_Q}\sum_{i\in\bbI_j}\frac{h_i^2}{q_i} \quad\forall q\in \ri(Q) \mbox{ and } \forall h\in\R^n.
\end{equation}
\end{lemma}
\begin{proof}
We will first show the following inductive hypothesis over the set of non-root simplexes 
  $\nonrootsimplexes =\left\{ j\in S_Q:~b_Q^j > 0 \right\}$
for any depth $d\geq 0$:
\begin{align*}
  \sum_{j\in {\nonrootsimplexes}:d_j \leq d} \beta_j \left[ \sum_{i\in \bbI_j}\left( \frac{h_i^2}{q_i}  -  \frac{2h_ih_{p_j}}{q_{p_j}} \right)   + \frac{h_{p_j}^2}{q_{p_j}}\right]
  - \sum_{j\in {\nonrootsimplexes}:d_j\leq d} \sum_{i\in \bbI_j} \frac{h_i^2}{q_i}
  \geq
  - \sum_{j\in {\nonrootsimplexes}:d_j=d} \frac{\beta_j\alpha_j}{\beta_j-\alpha_j}\frac{h_{p_j}^2}{q_{p_j}}
\end{align*}


We begin with the inductive step, as the base case will follow from the same logic. Consider a treeplex $Q$ of depth $d>0$. By applying the inductive hypothesis we have
\begin{align}
 & \sum_{j\in {\nonrootsimplexes}:d_j \leq d} \beta_j \left[ \sum_{i\in \bbI_j}\left( \frac{h_i^2}{q_i}  -  \frac{2h_ih_{p_j}}{q_{p_j}} \right)   + \frac{h_{p_j}^2}{q_{p_j}}\right]
   - \sum_{j\in {\nonrootsimplexes}:d_j \leq d} \sum_{i\in \bbI_j} \frac{h_i^2}{q_i} \nonumber\\
  \geq &
  \sum_{j\in {\nonrootsimplexes}:d_j = d} \beta_j \left[ \sum_{i\in \bbI_j}\left( \frac{h_i^2}{q_i}  -  \frac{2h_ih_{p_j}}{q_{p_j}} \right)   + \frac{h_{p_j}^2}{q_{p_j}}\right]
  - \sum_{j\in {\nonrootsimplexes}:d_j = d} \sum_{i\in \bbI_j} \frac{h_i^2}{q_i}
         - \sum_{j\in {\nonrootsimplexes}:d_j=d-1} \frac{\beta_j\alpha_j}{\beta_j-\alpha_j}\frac{h_{p_j}^2}{q_{p_j}} \label{eq:recurrence_after_induction}
\end{align}
Now we can rearrange terms: The sum over $j\in {\nonrootsimplexes}$ such that $d_j=d-1$ is  equivalent to a sum over the immediate descendant information sets $k\in\cD_j^i$ inside the square brackets, and we can move the sum over $i\in\bbI_j$ outside the square brackets by using the fact that $\sum_{i\in \bbI_j}\frac{q_i}{q_{p_j}} = 1$ and splitting the term $\frac{h_{p_j}^2}{q_{p_j}}$ into separate terms multiplied by $\frac{q_i}{q_{p_j}}$, this gives
\begin{align}
  \eqref{eq:recurrence_after_induction} = &
      \sum_{j\in {\nonrootsimplexes}:d_j = d}   \sum_{i\in \bbI_j} \left[ \left( \beta_j - 1 - \sum_{k\in \cD_j^i} \frac{\beta_k\alpha_k}{\beta_k - \alpha_k} \right)\frac{h_i^2}{q_i} - \left( \frac{2\beta_jh_ih_{p_j}}{q_{p_j}} \right)   + \frac{q_i\beta_jh_{p_j}^2}{q_{p_j}^2}\right] \nonumber\\
  \geq &
  \sum_{j\in {\nonrootsimplexes}:d_j = d}   \sum_{i\in \bbI_j} \left[ \left( \beta_j - \alpha_j \right)\frac{h_i^2}{q_i} - \left( \frac{2\beta_jh_ih_{p_j}}{q_{p_j}} \right)   + \frac{q_i\beta_jh_{p_j}^2}{q_{p_j}^2}\right], \label{eq:weight_increase}
\end{align}
 where the last inequality follows from the definition of $\alpha_j$. 

For indices $j\in S_Q$ such that $b^j_Q>0$ and $i\in \bbI_j$, the relations in
(\ref{eq:weight_recurrence}) imply $\beta_j > \alpha_j$, and so the expression
inside the square brackets in (\ref{eq:weight_increase}) is a convex function of
$h_i$. Taking its derivative {\wrt} $h_i$ and setting it to zero gives
$h_i=\frac{\beta_j}{\beta_j - \alpha_j}\frac{q_i}{q_{p_j}}h_{p_j}$. Thus, we
arrive at
\begin{align*}
  (\ref{eq:weight_increase}) \geq &
  \sum_{j\in {\nonrootsimplexes}:d_j = d}   \sum_{i\in \bbI_j} \left[ \frac{\beta_j^2}{\beta_j - \alpha_j}\frac{q_ih_{p_j}^2}{q_{p_j}^2} - \frac{\beta_j^2}{\beta_j - \alpha_j} \frac{2q_ih_{p_j}^2}{q_{p_j}^2} + \frac{q_i\beta_jh_{p_j}^2}{q_{p_j}^2}\right] \\
  = & \sum_{j\in {\nonrootsimplexes}:d_j = d} \frac{h_{p_j}^2}{q_{p_j}}\left[ \big(\frac{-\beta_j^2}{\beta_j - \alpha_j} + \beta_j\big) {\sum_{i\in \bbI_j} q_i \over q_{p_j}}\right]
      = - \sum_{j\in {\nonrootsimplexes}:d_j = d} \frac{\beta_j\alpha_j}{\beta_j - \alpha_j} \frac{h_{p_j}^2}{q_{p_j}}.  
\end{align*}
%
%
Hence, the induction step is complete. For the base case $d=0$ we do not need the inductive assumption: Because $\cD_j^i=\emptyset$, $\alpha_j= 1$, and we get \eqref{eq:weight_increase} by definition; we can then apply the same convexity argument. This proves our inductive hypothesis.

Then using Lemma~\ref{lem:hessian},  we now have
\begin{align*}
  h^\top\nabla^{2}\omega(q)h - \sum_{j\in S_Q}\sum_{i\in\bbI_j}\frac{h_i^2}{q_i} 
    &=  \sum_{j\in S_Q} \beta_j \left[ \sum_{i\in \bbI_j}\left( \frac{h_i^2}{q_i}  -  \frac{2h_ih_{p_j}}{q_{p_j}} \right)   + \frac{h_{p_j}^2}{q_{p_j}}\right] -  \sum_{j\in S_Q}\sum_{i\in\bbI_j}\frac{h_i^2}{q_i} \\
  &\geq  \sum_{j\in S_Q: b_Q^j = 0}  \left[ \sum_{i\in \bbI_j} \beta_j\frac{h_i^2}{q_i}  - \sum_{k \in \cD_j^i}  \frac{\beta_k\alpha_k}{\beta_k-\alpha_k}\frac{h_{i}^2}{q_i} - \frac{h_i^2}{q_i}\right]
 \geq 0.
\end{align*}
The first inequality follows from the fact that $h_{p_j}=0$ for all $j\in S_Q$ such that $b_Q^j=0$, and for all $j\in S_Q$ such that $b_Q^j>0$, we  used our induction. The last inequality follows from (\ref{eq:weight_recurrence}) and $q_i,h_i^2\geq0$. This then proves \eqref{eq:strong_convexity_rhs_sum}.
\end{proof}


We are now ready to prove our two main theorems, which we restate before proving
them. \strongconvexityltwo*
\begin{proof}
Since $q_i\leq 1$, Lemma~\ref{le:strong_convexity_rhs_sum} implies $h^\top\nabla^{2}\omega(q)h \geq \sum_{j\in S_Q}\sum_{i\in\bbI_j}h_i^2 =  \|h\|_2^2$ for all $q\in \ri(Q)$ and for all $h\in\R^n$. Because the dilated  entropy function $\omega(q)$ is twice differentiable on $\ri(Q)$, from  Fact~\ref{fac:strong_convexity_hessian}, we conclude that $\omega(\cdot)$ is   strongly convex {\wrt} the $\ell_2$ norm on $Q$ with modulus $1$. 
  
This analysis is tight: By choosing a vector $q\in \{0,1\}^{|Q|}$ such that $\|q\|_1=M_Q$,  and setting $h_i = \frac{\beta_j}{\beta_j - \alpha_j}\frac{q_i}{q_{p_j}}h_{p_j}$ for all indices $i$ such that $q_i=1$ and  $h_i=0$ otherwise, every inequality in the proof of Lemma~\ref{le:strong_convexity_rhs_sum} becomes an equality.
\end{proof}

\strongconvexitylone*
\begin{proof}
  To show strong convexity with modulus $1$ {\wrt} the $\ell_1$ norm, we lower bound
  the right-hand side of~\eqref{eq:strong_convexity_rhs_sum} in Lemma~\ref{le:strong_convexity_rhs_sum}:
\begin{align*}
  \sum_{j\in S_Q}\sum_{i \in \bbI_j} \frac{h_i^2}{q_i} 
  &\geq  \frac{1}{M_Q} \bigg(\sum_{j\in S_Q}\sum_{i \in \bbI_j}q_i\bigg) \sum_{j\in S_Q}\sum_{i \in \bbI_j} \frac{h_i^2}{q_i} 
  \geq  \frac{1}{M_Q}\bigg( \sum_{j\in S_Q}\sum_{i \in \bbI_j}   \frac{| h_i |}{\sqrt{q_i}} \sqrt{q_i} \bigg)^2 
  = \frac{1}{M_Q}  \| h \|_1^2, 
\end{align*}
where the first inequality follows from the fact that $M_Q$ is an upper bound on $\|q\|_1$ for any $q\in Q$, and the second inequality follows from the Cauchy-Schwarz inequality.

Hence, we deduce $h^\top\nabla^{2}\omega(q)h\geq\frac{1}{M_Q} \| h \|_1^2$ holds
for all $q\in \ri(Q)$ and for all $h\in\R^n$. Because the dilated entropy
function $\omega(q)$ is twice differentiable on $\ri(Q)$, from
Fact~\ref{fac:strong_convexity_hessian}, we conclude that $\omega(\cdot)$ is
strongly convex {\wrt} the $\ell_1$ norm on $Q$ with modulus
$\varphi=\frac{1}{M_Q}$.
\end{proof}

\subsection{Treeplex width} \label{sec:max_distance}
The convergence rates of {\fom}s such as {\mprox} and {\egt} algorithms  depend
on the diameter-to-strong convexity parameter ratio $\frac{\Omega}{\varphi}$, as
described in Section~\ref{sec:mirror_prox}. In order to establish full results
on the convergence rates of these {\fom}s, we now bound this ratio using
Corollary~\ref{co:strong_convexity_dilated_entropy} scaled by $M_Q$.


\begin{theorem}\label{the:entropy_diameter}
  For a treeplex $Q$, the dilated entropy function with simplex weights $\beta_j=M_Q(2+\sum_{r=1}^{d_j}2^{r}(M_{Q_j,r}-1))$ for each $j\in S_Q$ results in  
$
  \frac{\Omega}{\varphi} \leq M_Q^2 2^{d_Q+2}\log m
$ where $m$ is the dimension of the largest simplex $\Delta^j$ for $j\in S_Q$ in the treeplex structure. 
\end{theorem}


\section{\egt\ for extensive-form game solving}
\label{sec:mp_for_efg}
We now describe how to instantiate \egt\ for solving two-player zero-sum {\efg}s of the form \eqref{eq:sequence_form_objective} with treeplex domains. 
Below we state the customization of all the definitions from Section~\ref{sec:assumptions_and_setup} for our problem. 

Let $m$ be the size of the largest simplex in either of the treeplexes
$\cX,\cY$. Because $\cX$ and $\cY$ are treeplexes, it is immediately apparent
that they are closed, convex, and bounded. We use the $\ell_1$ norm on both of
the embedding spaces $\bE_x,\bE_y$. As our {\dgf}s for $\cX,\cY$ are compatible
with the $\ell_1$ norm, we use the dilated entropy \dgf\ scaled with weights
given in
Theorem~\ref{the:entropy_diameter}. 
Then Theorem~\ref{the:entropy_diameter} 
gives our bound on ${\Omega_\cX\over\varphi_\cX}$ and
${\Omega_\cY\over\varphi_\cY}$. Because the dual norm of the $\ell_1$ norm is
the $\ell_{\infty}$ norm, the matrix norm is given by: $
\|A\|=\max_{y\in \cY}\left\{ \| Ay \|_1^* :~ \|y\|_1=1\right\} = \max_{i,j}
|A_{i,j}|. $

\begin{remark}\label{rem:Lipschitz}
Note that $\|A\|$ is not at the scale of the maximum payoff difference in the original game. The values in $A$ are scaled by the probability of the observed nature outcomes on the path of each sequence.
Thus, $\|A\|$ is exponentially smaller (in the number of observed nature steps on the path to the maximizing sequence) than the maximum payoff difference in the original {\efg}.
\end{remark}

Theorem~\ref{the:entropy_diameter} immediately leads to the following convergence rate result for {\fom}s equipped with dilated entropy {\dgf}s to solve {\efg}s (and more generally {\bspp}s over treeplex domains). 
\begin{theorem}\label{thm:D-EntropyRate}
  Consider a {\bspp} over treeplex domains $\cX,\cY$. Then {\egt} algorithm equipped
  with the dilated entropy {\dgf} with weights $\beta_{j}=2+
  \sum_{r=1}^{d_j}2^{r}(M_{\cX_{j},r}-1)$ for all $j \in S_{\cX}$ and the corresponding setup for $\cY$ will return an $\e$-accurate solution to the {\bspp} in at most the following number of iterations:
\[\frac{\max_{i,j}|A_{i,j}|\, \sqrt{M_\cX^22^{d_\cX+2}M_\cY^22^{d_\cY+2}}\, \log m}{\epsilon}.\]
\end{theorem}


This rate in Theorem~\ref{thm:D-EntropyRate}, to our knowledge, establishes the state-of-the-art for {\fom}s with $O({1\over \e})$ convergence rate for {\efg}s.

\subsection{Improvements in extensive-form game convergence rate}
\label{sec:theoretical_improvement}

The ratio ${\Omega \over \varphi}$ of set diameter over the strong convexity parameter is important for FOMs that rely on a prox function, such as \egt\ and \mprox. 
Compared to the rate obtained by \citep{Kroer15:Faster}, we get the following improvement:
for simplicity, assume that the number of actions available at each information set is on average $a$, then our bound improves the convergence rate of \citep{Kroer15:Faster} by a factor of $\Omega(d_\cX \cdot a^{d_\cX} + d_\cY \cdot a^{d_\cY})$.

As mentioned previously, \citet{Hoda10:Smoothing} proved  only explicit bounds for the special case of uniform treeplexes 
that are constructed as follows: 1) A base treeplex $Q_b$ along with a subset of $b$ indices from it for branching operations is chosen. 2) At each depth $d$, a Cartesian product operation of size $k$ is applied. 3) Each element in a Cartesian product is an instance of the base treeplex with a size $b$ branching operation leading to depth $d-1$ uniform treeplexes constructed in the same way. 
Given bounds $\Omega_b,\varphi_b$ for the base treeplex, the bound of \citet{Hoda10:Smoothing} for a uniform treeplex with $d$ uniform treeplex levels (note that the total depth of the constructed treeplex is $d\cdot d_{Q_b}$, where $d_{Q_b}$ is the depth of the base treeplex $Q_b$) is
\[
\frac{\Omega}{\varphi}\leq O\left(b^{2{d}-2}k^{2{d}+2}{d}^2M_{Q_b}^2\frac{\Omega_b}{\varphi_b}\right).
\]
Then when the base treeplex is a simplex of dimension ${m}$, their bound for the dilated entropy on a uniform treeplex $Q$ becomes
\[
\frac{\Omega}{\varphi} \leq O\left(\left|S_{Q}\right|^2d_{Q}^2\log m \right).
\]
Even for the special case of a uniform treeplex with a base simplex, comparing Theorem~\ref{the:entropy_diameter} to their bound,
we see that our general bound improves the associated constants by exchanging
$O(\left|S_Q\right|^2d_{Q}^2)$ with $O(M_Q^22^{d_Q})$. Since $M_Q$ does not
depend on the branching operation in the treeplex, whereas $|S_Q|$ does,
these are also the first bounds to remove an exponential dependence on the
branching operation (we have only a logarithmic dependence). In
Example~\ref{ex:square_root_m} we showed that there exist games where $M_Q=\sqrt{|S_Q|}$, and in general $M_Q$ is much smaller than $|S_Q|$. Consequently, our results establish the best known convergence results for all {\fom}s based on dilated entropy {\dgf} such as \egt, \mprox, and stochastic variants of {\bspp} algorithms.

\cfr, \cfrp, and \egt\ all need to keep track of a constant number of current
and/or average iterates, so the memory usage of all three algorithms is of the
same order; when gradients are computed using an iterative approach as opposed
to storing matrices or matrix decompositions, each algorithm requires a constant
times the number of sequences in the sequence-form representation. Therefore, we
compare mainly the number of iterations required by each algorithm. Since the theoretical properties of {\cfr} and {\cfrp} are comparable, we compare to {\cfr}, with all statements being valid for {\cfrp} as well.

\cfr\ has a $O(\frac{1}{\e^2})$ convergence rate; but its dependence on the
number of information sets is only linear (and sometimes
sublinear~\citep{Lanctot09:Monte}). Since our results have a quadratic
dependence on $M_Q^2$, \cfr\ sometimes has a better dependence on game constants
and can be more attractive for obtaining low-quality solutions quickly for games
with many information sets. \mccfr\ and {\cfrp} have a similar convergence
rate~\citep{Lanctot09:Monte}, though {\mccfr} has cheaper iterations.

\citet{Gilpin12:First} give an equilibrium-finding algorithm presented as $O(\ln(\frac{1}{\e}))$; but this form of their bound has a dependence on a certain condition number of the $A$ matrix. Specifically, their iteration bound for sequential games is $O(\frac{ \| A \|_{2,2} \cdot \ln(\|A\|_{2,2} / \e) \cdot \sqrt{D}}{\delta(A)})$, where $\delta(A)$ is the condition number of $A$, $\|A\|_{2,2}=\sup_{x\ne 0} \frac{\|Ax\|_2}{\|x\|_2}$ is the Euclidean matrix norm, and $D=\max_{x,\bar{x}\in\cX,y,\bar{y}\in\cY} \| (x,y)-(\bar{x},\bar{y}) \|_2^2$.
Unfortunately, the condition number 
$\delta(A)$ is only shown to be finite for these games.
Without any such unknown quantities based on condition numbers, 
\citet{Gilpin12:First} establish a convergence rate of $O(\frac{\|A\|_{2,2}\cdot D}{\e})$. This algorithm, despite having the same dependence on $\epsilon$ as ours in its convergence rate, i.e., $O({1\over\epsilon})$, suffers from worse constants. In particular, there exist matrices such that $\|A\|_{2,2}=\sqrt{\|A\|_{1,\infty}\|A\|_{\infty,1}}$, where $\|A\|_{1,\infty}$ and $\|A\|_{\infty,1}$ correspond to 
the maximum absolute column and row sums, respectively.  Then together with the value of $D$, this leads to a cubic dependence on the dimension of $Q$. For games where the players have roughly equal-size strategy spaces, this is equivalent to a constant of $O(M_Q^4)$ as opposed to our constant of $O(M_Q^2)$.

\section{Numerical experiments} 
\label{sec:experiments}
We carry out numerical experiments to investigate the practical performance of
{\egt} on {\efg}s when instantiated with our {\dgf}. 

We test these algorithms on a scaled up variant of the poker game Leduc
holdem~\citep{Southey05:Bayes}, a benchmark problem in the imperfect-information
game-solving community. In our version, the deck consists of $k$ pairs of cards
$1\ldots k$, for a total deck size of $2k$. Each player initially pays one chip
to the pot, and is dealt a single private card. After a round of betting, a
community card is dealt face up. After a subsequent round of betting, if neither
player has folded, both players reveal their private cards. If either player
pairs their card with the community card they win the pot. Otherwise, the player
with the highest private card wins. In the event both players have the same
private card, they draw and split the pot.

First, we investigate the impact of applying the weights used in
recurrence~\eqref{eq:weight_recurrence}, as compared to the previous scheme
introduced in \citet{Kroer15:Faster}. To instantiate
recurrence~\eqref{eq:weight_recurrence} we have to choose a way to set $\beta_j$
relative to $\alpha_j$. Experimentally, we found that the best way to
instantiate the recurrence is to use $\beta_j=\alpha_j$ for all $j$, in spite of
the strict inequality required for our proof. This scheme will henceforth be
referred to as new weights. We compare these new weights to the weights used in
\citet{Kroer15:Faster} (henceforth referred to as old weights).
\begin{figure}[]
  \centering
    \includegraphics[width=0.49\textwidth]{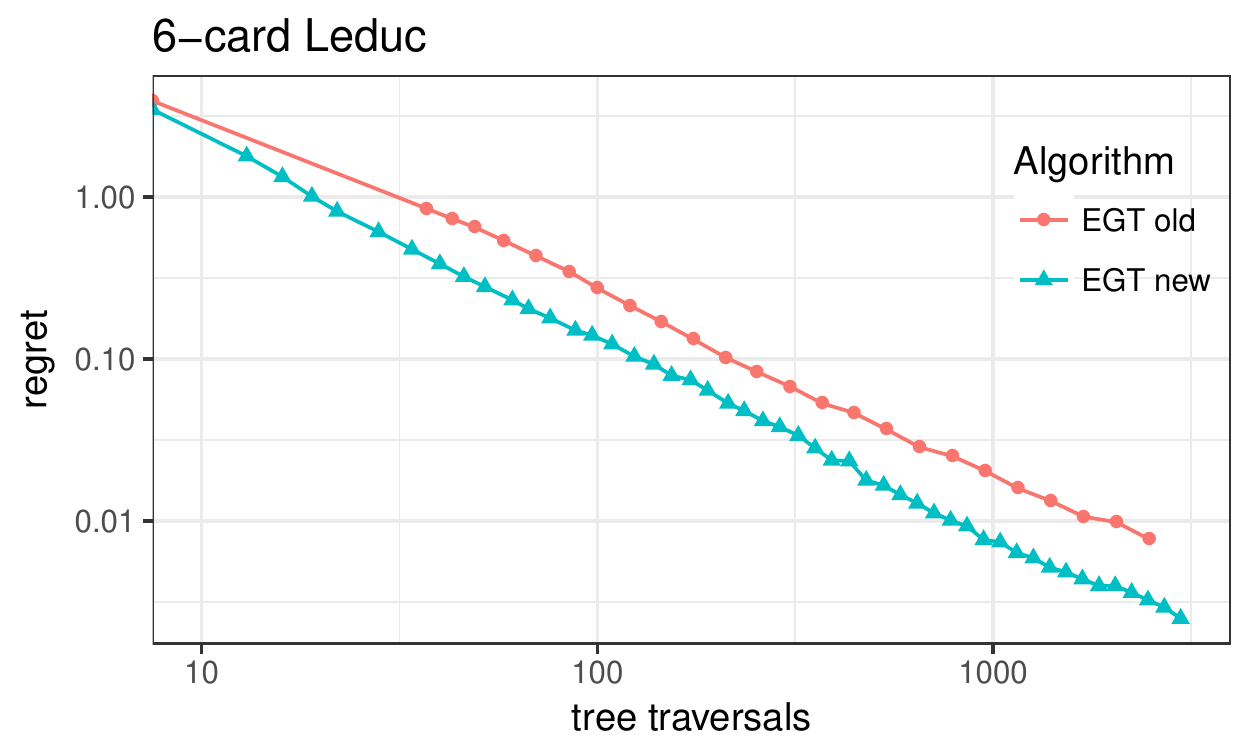}
    \includegraphics[width=0.49\textwidth]{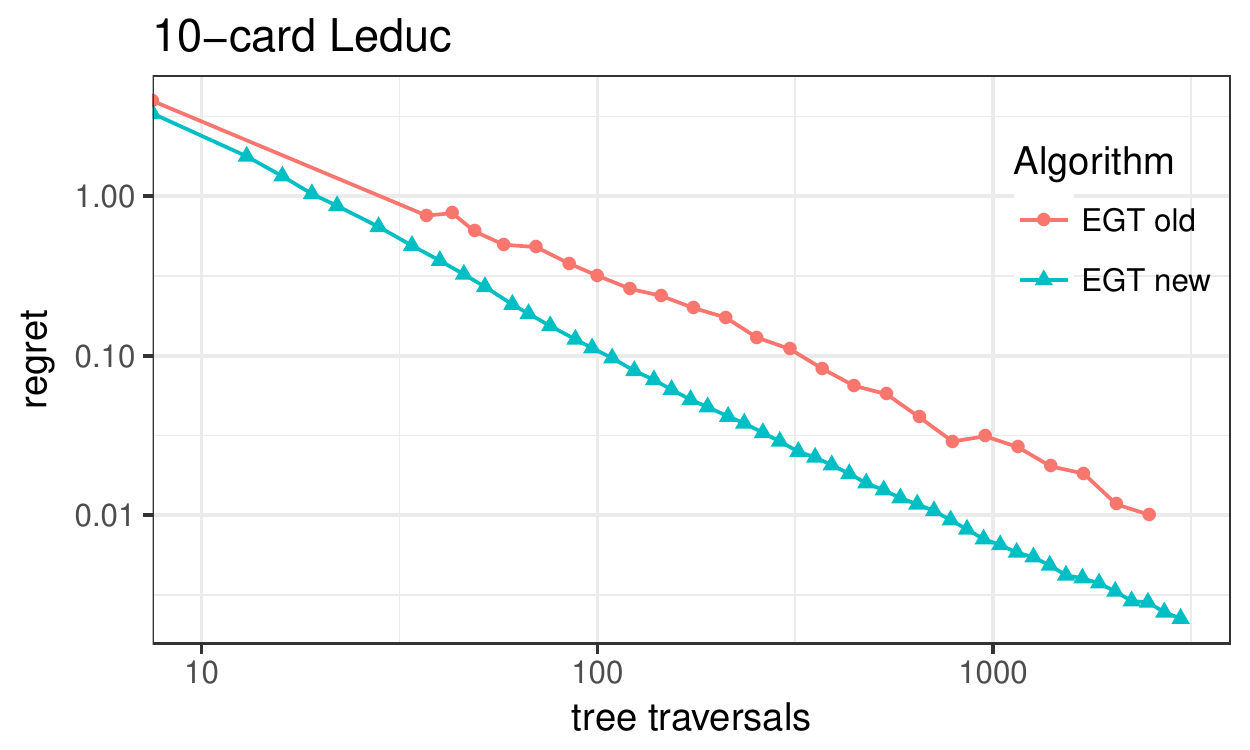}
    \includegraphics[width=0.49\textwidth]{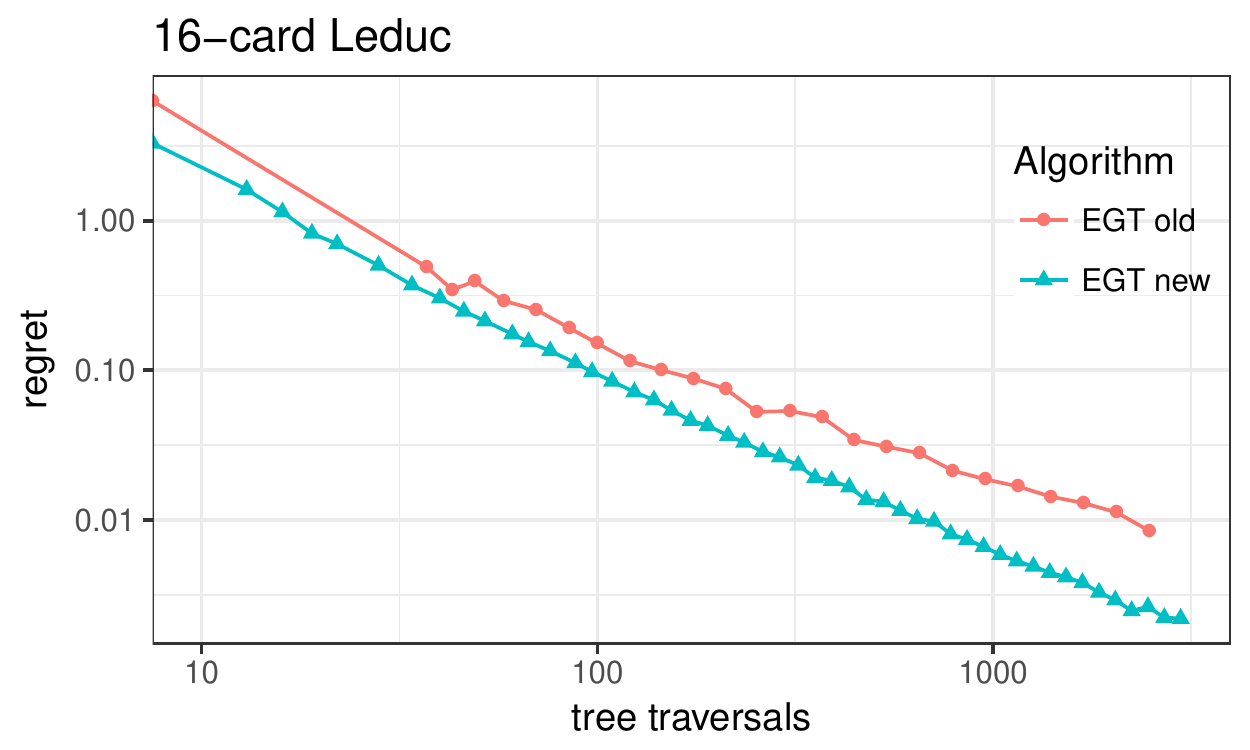}
    \includegraphics[width=0.49\textwidth]{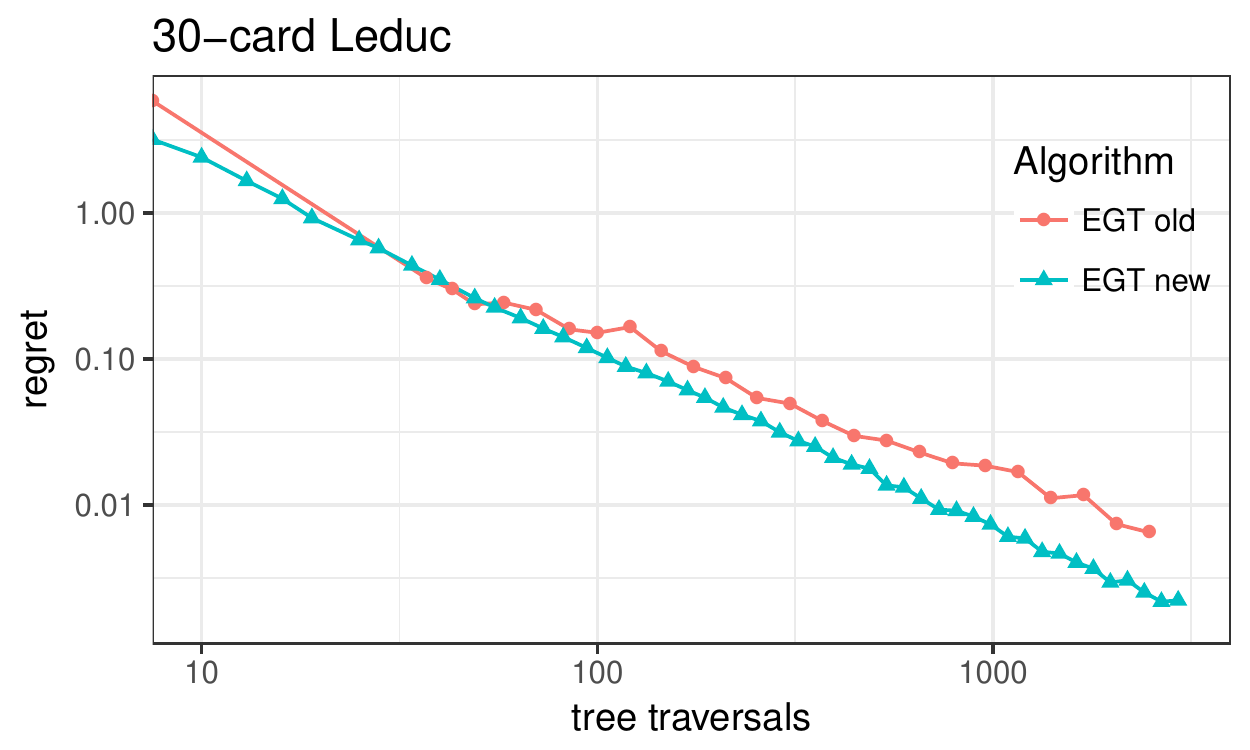}
    \caption{Regret as a function of the number of iterations for {\egt} with
      our weighting scheme (\egt\ new) and with the weighting scheme from
      \citet{Kroer15:Faster} (\egt\ old). Both axes are on a log scale.}
  \label{fig:experimental_results_better_bounds}
\end{figure}
Figure~\ref{fig:experimental_results_better_bounds} shows the result of running
\egt\ with the old and the new weights. For both the old and the new weights, we
found that the scalars $M_Q$ and $|S_Q|$ applied to each DGF in order to achieve
strong convexity modulus $1$ according to
Corollary~\ref{co:strong_convexity_dilated_entropy} and Theorem~$5.4$ of
\citet{Kroer15:Faster}, respectively, are too conservative. Instead, we show the
results after tuning these parameters for the corresponding algorithms to yield
the best results for each weight scheme. Anecdotally, we found that the old
weights are more sensitive and more difficult to tune. The performance also
seems more jittery; this is evident even in the strongest parameter we found
(especially noticeable on 10, 16, and 30-card Leduc in
Figure~\ref{fig:experimental_results_better_bounds}).

We compare the
performance of {\egt} to that of {\cfr} and {\cfrp} algorithms on
a scaled up variant of the poker game Leduc hold'em~\citep{Southey05:Bayes}, a
benchmark problem in the imperfect-information game-solving community. In our
version, the deck consists of $k$ pairs of cards $1\ldots k$, for a total deck
size of $2k$. Setting $k=3$ yields the standard Leduc game. Each player
initially pays one chip to the pot, and is dealt a single private card. After a
round of betting, a community card is dealt face up. After a subsequent round of
betting, if neither player has folded, both players reveal their private cards.
If either player pairs their card with the community card, they win the pot.
Otherwise, the player with the highest private card wins. In the event both
players have the same private card, they draw and split the pot.

\begin{figure}[]
  \centering
    \includegraphics[width=0.49\textwidth]{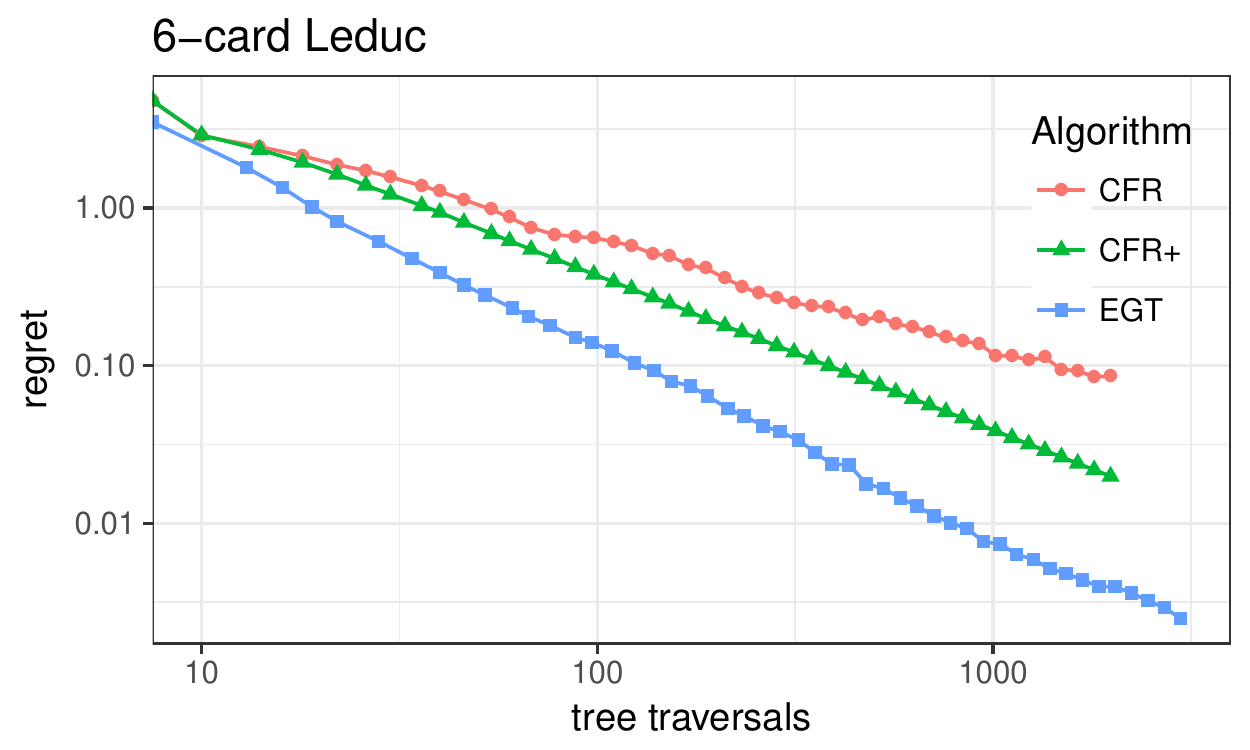}
    \includegraphics[width=0.49\textwidth]{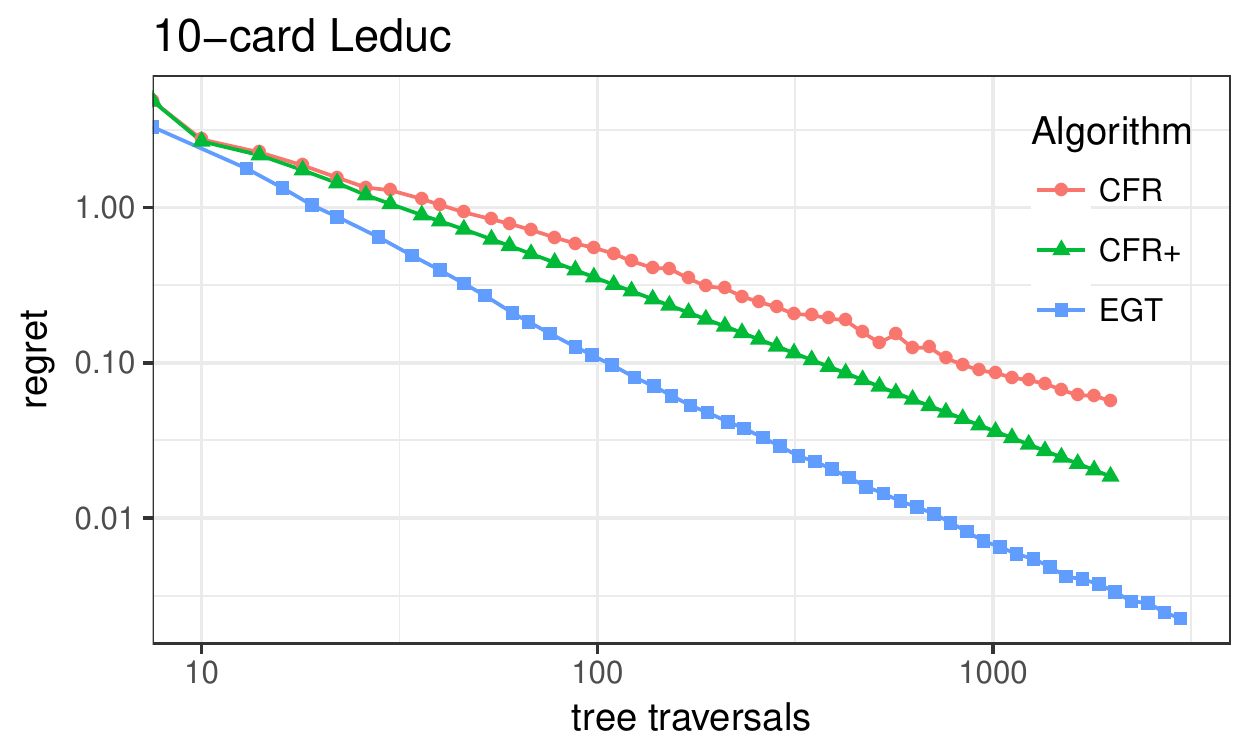}
    \includegraphics[width=0.49\textwidth]{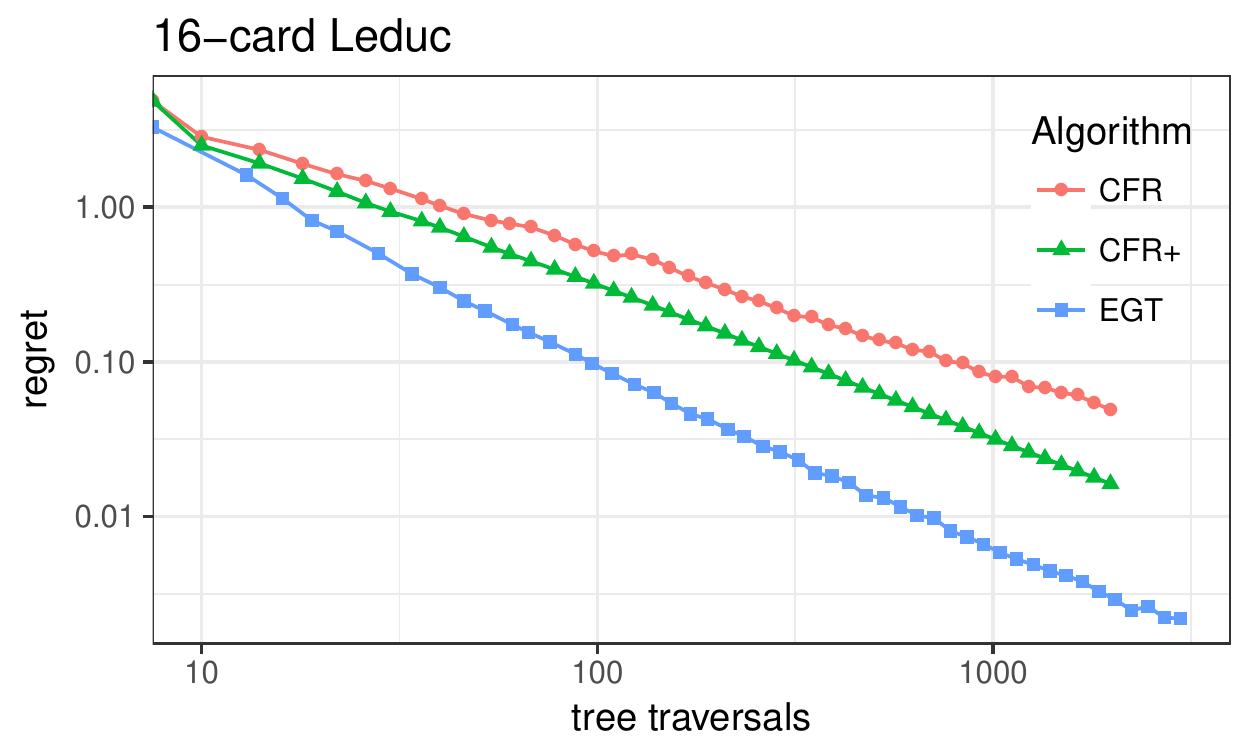}
    \includegraphics[width=0.49\textwidth]{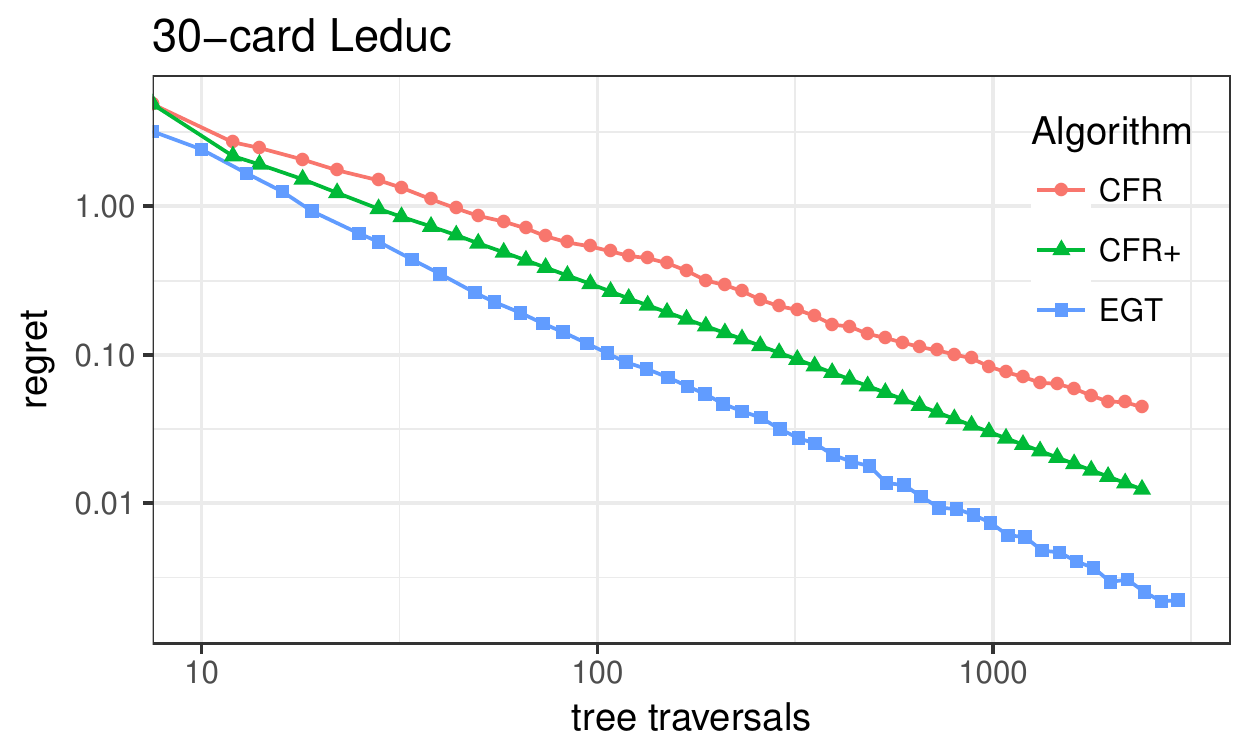}
  \caption{Regret as a function of the number of tree traversals in four different variants of Leduc hold'em for the \cfr, \cfrp, and \egt\ algorithms. Both axes are shown on a log scale.}
  \label{fig:experimental_results_all_algos}
\end{figure}

The results are shown in Figure~\ref{fig:experimental_results_all_algos}. Each
graph is a loglog plot that shows the results for a particular instance of Leduc
with $6,10,16$ and $30$ card decks, respectively. For each graph, we show the
performance of all three algorithms, with the x-axis showing the number of tree
traversals, and the y-axis showing the sum of regrets over the two players.
We note that tree-travels is a good proxy for overall computational effort because the majority of the time in FOMs is spent on gradient computations, which in our case directly translates into tree-traversals. 
We find that \egt\ instantiated with our \dgf\ significantly outperforms both \cfr\ and \cfrp\ across all four variants of Leduc. This is the case across all iterations; \egt\ finds a stronger initial point in $x^0,y^0$ (see Algorithm~\ref{alg:egt}), and maintains a stronger convergence rate across all iterations.

The performance we get from \egt\ relative to {\cfr} and {\cfrp} is surprising due
to what the conventional wisdom in the field has been. In \citet{Kroer15:Faster}
it was found that, while \egt\ has better convergence rate, {\cfr} (which performs
worse than {\cfrp}) had better initial performance, and it was only after a
certain number of iterations that \egt\ took over. Furthermore, the switch 
point where \egt\ is preferable was found to shift outward on the x-axis as the
Leduc game size was increased. This sentiment has been mirrored by
\citet{Brown16:Strategy-Based}. In contrast to this, we find that our \dgf\ along
with proper initialization leads to \egt\ performing better than not only {\cfr},
but also {\cfrp}, at every point on the x-axis. Furthermore, scaling up the game
size does not seem to adversely affect this relationship.

While the experiments in Figure~\ref{fig:experimental_results_all_algos} are
very interesting from the perspective of which algorithm to use for large-scale
{\efg}-solving in practice going forward, there are some caveats to keep in mind. First, we only considered number of tree traversals in our performance calculations. However, {\cfr} algorithms have the ability to avoid parts of the tree traversal. For games where accelerated best-response
calculation~\citep{Johanson11:Accelerating} can be applied, e.g., poker-like
games, this is unlikely to have a big effect. But, for some other games, this
aspect can be important, though note that \citet{Brown17:Dynamic} showed
experimentally that pruning can be used in \egt\ as well. 
Second, to get superior performance from \egt, we had to hand-tune initialization parameters relating to our \dgf, whereas {\cfrp} requires no tuning. Development of an algorithmic scheme for choosing this tuning parameter in \egt\ can make it significantly easier to apply the tuned variant of \egt\ in practice. 
Third, on another practical aspect, {\cfrp} is a conceptually very simple algorithm, and thus also easy to implement. In contrast to this, \egt\ and our \dgf\ requires a safe-guarded numerical implementation because the prox operator associated with our \dgf\  requires taking exponentials.

\section{Conclusions}\label{sec:conclusions}
We have investigated {\fom}s for computing Nash equilibria in two-player
zero-sum perfect-recall {\efg}s. On the theoretical side, we analyzed the strong
convexity properties of the dilated entropy \dgf\ over treeplexes. By
introducing specific weights that are tied to the structure of the treeplex, we
improved prior results on treeplex diameter from $O(|S_Q|M_Qd2^d\log{m})$ to
$O(M_Q^22^{d_Q+2}\log{m})$, thereby removing all but a logarithmic dependence on
branching associated with the branching operator in the treeplex definition. 
These results lead to significant
improvements in the convergence rates of many {\fom}s that can be equipped with
dilated entropy {\dgf}s and used for {\efg} solving including but not limited to
\egt, \mprox, and Stochastic \mprox.

We numerically investigated the performance of {\egt} and compared it to the
practical state-of-the-art algorithms {\cfr} and {\cfrp}. Our experiments showed
that \egt\ with the dilated entropy \dgf, when tuned with a proper scaling, has better practical, as well as theoretical, convergence rate than {\cfrp}, the current state-of-the-art algorithm in practice. While our scaling parameter for the {\dgf} did not require extensive tuning, we believe a more principled way of setting it is worthy of further future investigation.

Theorems~\ref{th:recurrence_bound_l2} and~\ref{th:recurrence_bound_l1} establish bounds for a general class of weights $\beta_j$ satisfying the recurrence~\eqref{eq:weight_recurrence}. Then in Corollary~\ref{co:strong_convexity_dilated_entropy}, we have selected a particular weighting scheme for $\beta_j$ satisfying \eqref{eq:weight_recurrence} and performed our numerical tests. There may be other interesting choices of $\beta_j$ satisfying the recurrence~\eqref{eq:weight_recurrence}. Thus, finding a way to optimally choose among the set of weights satisfying \eqref{eq:weight_recurrence} to minimize the polytope diameter for specific games is appealing.

On a separate note, in practice \cfr\ is often paired with an abstraction
technique~\citep{Sandholm10:State} such as those mentioned in
Section~\ref{sec:related_work}. This is despite the lack of any theoretical
justification. Effective ways to pair {\fom}s such as \mprox\ and \egt\ with
practical abstraction techniques~\citep{Brown15:Hierarchical} or abstraction
techniques that achieve solution-quality
guarantees~\citep{Lanctot12:no-regret,Kroer14:Extensive-Form,Kroer16:Extensive_Imperfect}
are also worth further consideration.


\bibliographystyle{plainnat}
\bibliography{dairefs.bib}
\clearpage
\appendix

\section{Omitted proofs}
\subsection{Proof of Lemma~\ref{lem:hessian}}
\begin{proof}
Consider $q\in \ri(Q)$ and any $h\in\R^n$. For each $j\in S_Q$ and $i\in\bbI_j$, the second-order partial derivates of $\omega(\cdot)$ {\wrt} $q_i$ are:
  \begin{equation}
    \nabla_{q_i^2}^2 \omega(q)=\frac{\beta_j}{q_i} + \sum_{k \in \cD^i_j}\sum_{l\in \bbI_{k}}\frac{\beta_{k}q_{l}}{q_i^2}= \frac{\beta_j}{q_i} + \sum_{k \in \cD^i_j} \frac{\beta_{k}}{q_i}, \label{eq:partial_derivative_diagonal}
  \end{equation}
  where the last equality holds because $k\in\cD^i_j$ and thus $\sum_{l\in \bbI_{k}} q_l = \|q^k\|_1=q_{p_k}=q_i$.  
Also, for each $j\in S_Q,i\in\bbI_j, k\in \cD^i_j$, and $l\in\bbI_k$, the second-order partial derivates {\wrt} $q_i,q_l$ are given by:
  \begin{equation}
    \nabla_{q_i,q_{l}}^2 \omega(q)= \nabla_{q_l,q_{i}}^2 \omega(q)
    = -\frac{\beta_{k}}{q_i}.
    \label{eq:partial_derivative_off_diagonal}
  \end{equation}
  Then equations~\eqref{eq:partial_derivative_diagonal} and~\eqref{eq:partial_derivative_off_diagonal} together imply
  \begin{align}
     h^\top\nabla^2\omega(q)h 
 =&  \sum_{j\in S_Q}\sum_{i\in \bbI_j} \left[ h_i^2 \left( \frac{\beta_j}{q_i} + \sum_{k \in \cD^i_j} \frac{\beta_{k}}{q_i} \right) -  \sum_{k \in \cD^i_j}\sum_{l \in \bbI_{k}} h_ih_{l}\frac{2\beta_{k}}{q_i}   \right] . \label{eq:hessian_quadratic_unsimplified} 
  \end{align}
Given $j\in S_Q$ and $i\in\bbI_j$, we have $p_k=i$ for each $k\in \cD_j^i$ and for any $k\in \cD_j^i$, there exists some other $j'\in S_Q$ corresponding to $k$  in the outermost summation. Then we can rearrange the following terms:
  \[
    \sum_{j\in S_Q}\sum_{i\in \bbI_j} h_i^2 \sum_{k \in \cD^i_j} \frac{\beta_{k}}{q_i}
    = \sum_{j\in S_Q} \beta_j \frac{h_{p_j}^2}{q_{p_j}} 
    ~\mbox{ and }~ \sum_{j\in S_Q}\sum_{i\in \bbI_j}  \sum_{k \in \cD^i_j}\sum_{l \in \bbI_{k}} h_ih_{l}\frac{2\beta_{k}}{q_i}
    = \sum_{j\in S_Q}\sum_{i\in \bbI_j}\beta_j\frac{2h_ih_{p_j}}{q_{p_j}}.
  \]
Using these two equalities in \eqref{eq:hessian_quadratic_unsimplified} leads to \eqref{eq:hessian_quadratic_count_at_parent} and proves the lemma.
\end{proof}

\subsection{Proof of Theorem~\ref{the:entropy_diameter}}
\begin{proof}
For our choice of scaled weights $\beta_j$, Corollary~\ref{co:strong_convexity_dilated_entropy} implies that the resulting dilated entropy function is strongly convex with modulus $\varphi=1$. Hence, we only need to bound $\Omega$. 

Any vector $q\in Q$ satisfying $q_i\in\{0,1\}$ for all $i$ maximizes $\omega(q)$
and results in $\max_{q\in Q}\omega(q)=0$. For the minimum value, consider any
$q\in\ri(Q)$. Applying the well-known lower bound of $-\log m$ for the negative entropy function on an $m$-dimensional simplex, we have
  \begin{align}
    \omega(q) & = \sum_{j\in S_Q} \beta_j q_{p_j}\sum_{i\in \bbI_j} \frac{q_i}{q_{p_j}}\log \frac{q_i}{q_{p_j}} 
    \geq -\sum_{j\in S_Q} \beta_j q_{p_j} \log m
    =-\sum_{d=0}^{d_Q} \sum_{j\in S_Q:d_j=d} \beta_j q_{p_j} \log m \nonumber\\
    & =-\sum_{d=1}^{d_Q} \sum_{j\in S_Q:d_j=d} \beta_j q_{p_j} \log m -  \sum_{j\in S_Q:d_j=0} \beta_j q_{p_j} \log m \nonumber\\
    & = - M_Q \log m \sum_{d=1}^{d_Q} \sum_{j\in S_Q:d_j=d} q_{p_j}\bigg(2+\sum_{r=1}^{d}2^{r}(M_{Q_j,r}-1)\bigg) -  M_Q\sum_{j\in S_Q:d_j=0} 2 q_{p_j} \log m \nonumber\\
    & \geq - M_Q \log m \sum_{d=1}^{d_Q} \sum_{j\in S_Q:d_j=d}q_{p_j} M_{Q_j}\sum_{r=1}^{d}2^{r} -  2M_Q \log m\sum_{j\in S_Q:d_j=0} q_{p_j}, \label{eq:treeplex_width1}
 \end{align}
where the last inequality follows because for each $j\in S_Q$ with $d_j=0$, the definition of $M_Q$ implies $\sum_{j\in S_Q:d_j=0}q_{p_j}\leq M_Q$, and
for each $j\in S_Q$ with $d_j=d\geq 1$, we have $2+\sum_{r=1}^{d}2^{r}(M_{Q_j,r}-1) \leq \sum_{r=1}^{d}2^{r}M_{Q_j,r} \leq \sum_{r=1}^{d}2^{r}M_{Q_j}$ since $M_{Q_{j,r}}\leq M_{Q_j}$. Also, from Fact~\ref{fact:M_Q}(b), we have $\sum_{j\in S_Q:d_j=d}q_{p_j} M_{Q_j} \leq M_Q$.
Then we arrive at
  \begin{align*}
    (\ref{eq:treeplex_width1}) 
    &\geq - M_Q^2 \log m \bigg(2+\sum_{d=1}^{d_Q} \sum_{r=1}^{d}2^{r} \bigg)
    = - M_Q^2 \log m \bigg(2+\sum_{d=1}^{d_Q} (2^{d+1}-2) \bigg) \\
    &= - M_Q^2 \log m \bigg(2+\sum_{d=1}^{d_Q} 2^{d+1} - 2d_Q \bigg)
    \geq - M_Q^2 (\log m) 2^{d_{Q}+2},  
 \end{align*}
where the last inequality follows because for $d_Q=0$ we have $2^{d_Q+2}=4>2$ and for $d_Q\geq 1$ we have $2d_Q\geq 2$.
 
This lower bound on the minimum value, i.e., $\min_{q\in Q}\omega(q)\geq - M_Q^2 (\log m) 2^{d_{Q}+2}$, coupled with $\max_{q\in Q}\omega(q)\leq 0$, establishes the theorem.
\end{proof}

\end{document}